\documentclass{aa}  
\usepackage{graphicx}
\usepackage{txfonts}
\usepackage{color}
\usepackage{multirow}
\usepackage[nice,loose]{units}            % Añadido por Pipi
\definecolor{orange}{rgb}{0.8, 0.33, 0.0}

\begin{document}

   \title{Constant entropy hybrid stars: a first approximation of cooling evolution}

%   \subtitle{I. Overviewing the $\kappa$-mechanism}

   \author{M. Mariani
          \inst{1, 2}
          \fnmsep\thanks{mmariani@fcaglp.unlp.edu.ar}
          \and
          M. Orsaria\inst{1, 2, 3}
          \fnmsep\thanks{morsaria@fcaglp.unlp.edu.ar}
          \and
          H. Vucetich\inst{1}
          \fnmsep\thanks{vucetich@fcaglp.unlp.edu.ar}
          }

   \institute{Grupo de Gravitación, Astrofísica y Cosmología,\\
   Facultad de Ciencias Astronómicas y Geofísicas,\\
             Paseo del Bosque S/N (1900), La Plata, Argentina\\
%              \email{wuchterl@amok.ast.univie.ac.at}
         \and
             Consejo Nacional de Investigaciones Científicas y Técnicas (CONICET),\\
             Rivadavia 1917, 1033 Buenos Aires, Argentina\\
%             \email{c.ptolemy@hipparch.uheaven.space}
%             \thanks{The university of heaven temporarily does not
%                     accept e-mails}
         \and
             Department of Physics, San Diego State University,\\
			5500 Campanile Drive, San Diego, CA 92182, USA}
%             \email{c.ptolemy@hipparch.uheaven.space}
%             \thanks{The university of heaven temporarily does not
%                     accept e-mails}
%             }

%   \date{Received September 15, 1996; accepted March 16, 1997}

% \abstract{}{}{}{}{} 
% 5 {} token are mandatory
 
\abstract
 {}
{We aim to study the possibility of a hadron-quark phase transition in the interior of neutron stars, taking into account different schematic evolutionary stages at finite temperature. We also discuss the strange quark matter stability in the quark matter phase. Furthermore, we aim to analyze the astrophysical properties of hot and cold hybrid stars, considering the constraint on maximum mass given by the pulsars  J1614-2230 and  J0348+0432.}
{We have developed a computational code to construct semi-analytical hybrid equations of state at fixed entropy per baryon and to  obtain different families of hybrid stars. An analytical approximation of the Field Correlator Method is developed for the quark matter equation of state. For the hadronic equation of state we use a table based on the relativistic mean field theory, without hyperons. The phase transition was obtained imposing the Maxwell conditions, by assuming a high surface tension at the interface hadron-quark. We solved the relativistic structure equations of hydrostatic equilibrium and mass conservation for hybrid star configurations.}
{For the different equations of state obtained, we calculated the stability window for the strange quark matter, lepton abundances, temperature profiles and contours profiles for the maximum mass star depending on the parameters of the Field Correlator Method. We also computed the mass-radius and gravitational mass-baryonic mass relationships for different hybrid star families.}
{We have analyzed different stages of hot hybrid stars as a first approximation of the cooling evolution of neutron stars with quark matter cores. We obtain cold hybrid stars with maximum masses $\geq 2 M_\odot$ for different combinations of the Field Correlator Method parameters. In addition, our study based on the gravitational mass - baryonic mass plane shows a late phase transition between hadronic and quark matter during the proto-hybrid star evolution, in contrast with previous studies of proto-neutron stars.}


\maketitle

%==========================================================================================================

\section{Introduction}

A proto-neutron star (proto-NS) is the remaining old degenerated core of an intermediate mass star (between 10 and 25 solar masses at the zero age main sequence) after a supernova type-II or type-Ib/Ic explosion \citep{Woosley:2002}. Its evolution could be described roughly by three characteristic isentropic stages: first, the new born neutron star (NS) has an entropy per baryon $s \sim 1$ (in units of Boltzmann constant), with an abundance of trapped neutrinos, $Y_{\nu} \neq 0$, whose fraction is set by requiring  the total fraction of leptons $Y_{\it l} \simeq 0.4$. This stage is dominated by neutrino diffusion, but as the neutrino mean free path is much smaller than the star radius, they remain trapped. The first $\sim 15$ seconds, the proto-NS radiates its lepton excess (deleptonization) and the neutrino flux produces an increase in the temperature. The star reaches the maximum heating in this second stage, and its inner matter has entropy per baryon $s \simeq 2$. Next, the neutrino mean free path begins to increase, becoming much larger than the radius of the star. The fraction $Y_{\nu}\neq 0$ evolves to a state in which $Y_{\nu} = 0$, because the matter becomes transparent to neutrinos.  Once the electron chemical potential reaches the muon mass value, muons start to appear. The arising of muons reduces the number of electrons and affects the proton fraction of the stellar matter. The deleptonization process continues, and after a few minutes the third stage takes place. The entropy per baryon decreases to $s = 0$ and the proto-star cools down remaining transparent to neutrinos with $Y_{\nu} = 0$. The resulting object after these three stages is a cold and stable NS  \citep{Burrows:1986,Steiner:2000}.

In the chaotic rearrangement period of deleptonization and cooling described above, a phase transition to an exotic state of matter, like hyperons, Bose condensates or quark matter, could appear \citep{Benvenuto:1999,Benvenuto:1989tf,Benvenuto:1989qr, Prakash:2001} inside the proto-NS. In the case of quark matter, studies of the Quantum Chromodynamics (QCD) phase diagram predicts that the crossover shown by Lattice calculations at vanishing chemical potential, will become a first-order phase transition at intermediate temperatures and high baryon chemical potentials (see \cite{Yin:2012} and references therein). It may be plausible that this scenario of matter under extreme conditions occurs in the interior of neutron stars, as we get closer to the innermost core of the star, from the outer to the inner core, where the matter is compressed to densities several times the nuclear matter saturation density ($n_0 \sim \unit[0.16] {fm^{-3}}$).

Most of the mass of a NS is contained in the inner and outer cores and its equation of state (EoS) has not yet been well determined. The discovery of the massive PSR J1614-2230 and PSR J0348+0432 ($\sim 2 M_{\odot}$) neutron stars, whose masses have been very accurately determined ($ \pm 0.04 M_{\odot}$) \citep{Demorest:2010,Antoniadis:2013}, has placed limits on a wide variety of EoS. Many EoS had to be ruled out, and the internal composition of NS had to be reconsidered, because a viable EoS should, at least, reproduce the mass value of these observed NS. 

At the end of the process of the NS formation, the degenerate neutron pressure partially counteracts the gravity force. The repulsive nuclear force beats the neutron degeneracy pressure resulting in less compact and more rigid structures supported by a stiffer EoS. For this reason neutron star masses are larger than  $\sim \unit[0.7] {M_{\odot}}$, the maximum star mass derived by Tolman-Oppenheimer and Volkoff equations \citep{tov1, tov2} if a non-interacting degenerated neutron gas is considered as forming neutron stars. Thus, strong interactions are crucial in obtaining neutron star configurations that are compatible with measured neutron star masses.

Some alternative models for the EoS include strange matter in form of hyperons (see, for example \cite{Bednarek:2011gd,Yamamoto:2014jga,Katayama:2015}). Although hyperons soften the EoS, massive NS with hyperons would be possible, considering certain relationships between the coupling constants and interactions between the particles involved in the hadronic matter composing their interiors \citep{Oertel:2016}. Another possibility is that NS could contain pure quark matter. However, to obtain stars as heavy as two solar masses, it should be feasible to adjust some of the parameters which mediates the interaction between quarks \citep{Orsaria:2013hna, Weber:2014qoa} and/or that quark matter is a color superconductor \citep{Bonanno:2012,Zdunik:2013}. Whatever the matter composing the interior of a NS may be (quarks or hadrons or a mixing of both phases), the equation of state describing the matter should be stiff enough to obtain massive NS.

Regarding quark matter inside compact objects, for many years the models most used to describe it have been the MIT bag model \citep{Chodos:1974je} and the Nambu Jona-Lasinio model \citep{Nambu:1961a,Nambu:1961b}. Some modifications to these models have been implemented to obtain a more suitable description of quark matter at high densities in the context of hybrid stars (HS), NS with a quark matter core surrounded by a shell of hadronic matter (see \cite{Chatterjee:2016} and references therein). Recently, for the description of quark matter phase inside cold NS, the Field Correlator Method (FCM) has been used \citep{Plumari:2013,Logoteta:2013,Burgio:2016}. The FCM is a non-perturbative approximation of QCD which includes, from the first
principles, the confinement dynamics in terms of field correlators. The model is parametrized through the gluonic condensate, $G_2$, and the quark-antiquark static potential for long distances, $V_1$, taking into account the confinement. These two parameters control the EoS of the deconfined quark phase.

In the framework of the isentropic stages described at the beginning, we have studied the occurrence of a first order phase transition from hadronic to quark matter in proto-NS, assuming the surface tension at the interface hadron-quark is such that a sharp phase transition is the favorable scenario for the deconfinement (\cite{Alford:2015, Ranea:2015ldr}). The hybrid EoS that models the proto-NS is obtained using the simplest method for constructing a first order phase transition by considering an isobaric transition with one conserved charge, for example, the baryonic chemical potential. This is known as Maxwell construction.

To obtain the phase transition and the astrophysical quantities for the study of hybrid stars, a computational code called NeStOR (Neutron-Star-Object Research) has been developed. This code contains a set of routines and scripts to calculate several EoS at finite temperature and constant entropy. The combination of different programming languages provided in the code establishes the essential conditions for an unique determination of the matter equilibrium composition in the proto-hybrid star, that is, the requirements of beta equilibrium, charge neutrality and the conservation of baryonic and leptonic number. The program also computes particle abundances and the stellar structure of a spherically symmetric compact object by integrating the relativistic equations of hydrostatic equilibrium for different model parameters. Through this code it is also possible to compare the values of the maximum baryonic mass of each family of stars for the three stages considered. Supposing that the baryonic mass remains constant during the process of thermal evolution, the gravitational mass of the star, which corresponds to the total stellar energy, changes during the evolution of the proto-NS. Then, a study of the evolution in a gravitational mass - baryonic mass plane must be taken into account, giving the possibility of knowing in which stage takes place the phase transition from hadronic matter to quark matter. The description of proto-NS evolution along constant baryonic mass sequences have been already discussed in \cite{Bombaci:1996}.

Thereby, using the NeStOR code, it is possible to analyze the thermal evolution of proto-hybrid stars through snapshot sequences at constant entropy, also considering their characteristic composition at each stage. A semi-analytical calculation of the FCM at finite temperature is used to describe the quark matter phase at the inner cores. For the hadronic phase, we employ a table containing an EoS created for astrophysical simulations of proto-NS \citep{Shen:2010a, Shen:2011, Shen:2010b}.

The paper is organized as follows. In Section \ref{section eos quark}, we give the thermodynamic expressions for the quark matter as
functions of temperature and chemical potential, and we discuss the choice of the parameters of the FCM as well as the stability of strange quark matter. The thermodynamic conditions for the construction of the hybrid EoS and the phase transition hadron-quark are presented in Section \ref{phase}, together with the description of the hadronic and leptonic EoS. In Section \ref{entop} we explain briefly the calculations at constant entropy, giving the snapshot sequences of thermal evolution. In Section \ref{hybrid} we show the results about the structure of hot and cold hybrid stars and we analyze the gravitational mass-baryonic mass plane for some representative combinations of the FCM parameters. A summary of the work and conclusions are provided in Section \ref{conclus}. Details of the calculations and treatment of quark matter are given in the Appendix.

%==========================================================================================================

\section{Quark matter EoS within the Field Correlator Method}
\label{section eos quark}
\subsection{Correlators and the phase diagram}

The microscopic EoS describing the matter in the inner core of the hot and cold hybrid stars (HHS and CHS henceforth) is based on the Field Correlator Method approach \citep{Dosch:1987,Dosch:1988,Simonov:1988}. In the FCM approach, it is assumed, as a first approximation, that single quark and gluon lines (single line approximation) interact in non-perturbative way, using strength field correlators, with vacuum fields (gluon condensates and quark condensates) \citep{Simonov:2007b}.

The confined and deconfined phases are governed by four Gaussian field correlators: the color-electric, $D^E(x)$, $D^E_1(x)$, and color-magnetic, $D^H(x)$, $D^H_1(x)$, field correlators . The fields $D^E$ and $D^H$ are of pure non-perturbative origin, whereas $D^E_1$ and $D^H_1$ contain all perturbative diagrams, but also non-perturbative contributions. The non-perturbative information is given through the string tension, quarks and gluons are connected by straight-line Nambu-Goto strings \citep{Nefediev:2009}, which is a characteristic integral of Gaussian correlators \citep{ DiGiacomo:2002}. Thus, in the lowest non-perturbative approximation the partition function factorizes into a product of one-gluon and one-quark (antiquark) contributions, and then the corresponding thermodynamic potential can be calculated.

\begin{figure}[h]
  \centering
  \includegraphics[width=0.95\columnwidth]{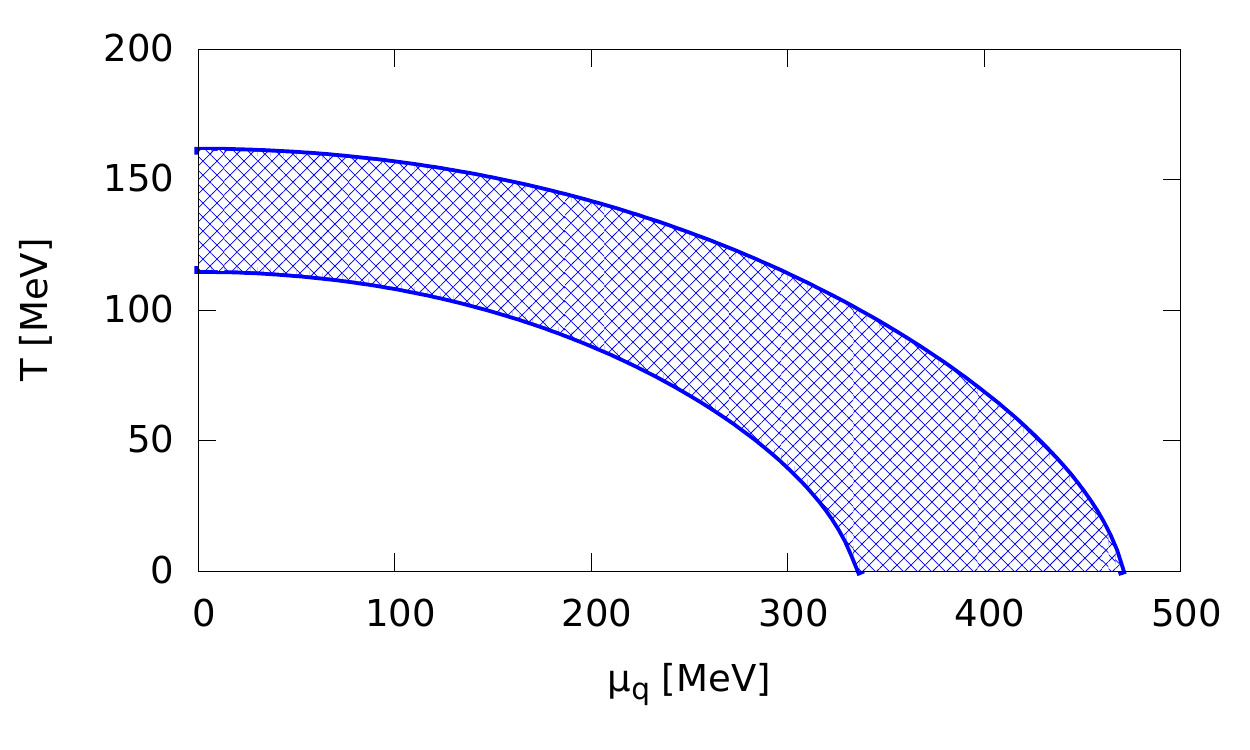}
  \caption{Schematic phase diagram in the $T-\mu$ plane. The lower curve defines the phase transition between confined and deconfined matter for $V_1=\unit[0.01] {GeV}$ and $G_2 = \unit[0.006] {GeV^4}$, while the upper one defines the phase transition curve for $V_1=\unit[0.10] {GeV}$ and $G_2 = \unit[0.016] {GeV^4}$. The shaded area indicates the region of the FCM parameters considered in this work, $\unit[0.01] {GeV} \le V_1 \le \unit[0.10] {GeV}$ and $\unit[0.006] {GeV^4} \le G_2 \le \unit[0.016] {GeV^4}$.}
  \label{phase_diag}%
\end{figure}

Through the thermodynamic potential, thermodynamic properties of the quark matter can be stated in terms of a $T-\mu$ phase diagram, as shown in Fig. (\ref{phase_diag}), where the confined phase (hadrons) is separated from the deconfined quark-gluon plasma phase. For temperatures $T > T_c$, being $T_c$ the critical temperature, the correlator $D^E(x)$, and the electric string tension associated, vanishes, while the deconfining electric part, $D^E_1(x)$, and magnetic correlators $D^H(x)$, $D^H_1(x)$, are the responsible for the dynamics of the deconfined phase.

The phase diagram shown in Fig. \ref{phase_diag}, obtained through the semi-analytical treatment detailed in Section \ref{therm_quant}, is in agreement with the results of \cite{Simonov:2007a, Simonov:2007b}.

\subsection{Thermodynamic quantities of the quark matter}
\label{therm_quant}

The FCM is parametrized through the gluon-condensate $G_2$ ,generated by $D^E(0)$, $D^E_1(0)$, $D^H(0)$ and $D^H_1(0)$, and the large distance static quark-antiquark, $q\bar q$, potential $V_1$, generated by $D^E_1(x)$, \citep{Simonov:2007b}. Even though the single line approximation turn off the quark-gluon interaction, a strong interaction of quarks and gluons with the non-perturbative vacuum field exist \citep{Nefediev:2009}.  In this way, using single-gluon and single-quark contributions, the quark and gluon pressures, $P_q$ and $P_g$, respectively, are given by \citep{Simonov:2007a, Simonov:2007b, Nefediev:2009}

\begin{equation}\label{presionq}
 P_q  = \frac{T^4}{\pi^2}\phi_\nu (\frac{\mu_q - V_1/2}{T}),
\end{equation}
where 
\begin{equation}
\phi_\nu (a) = \int_0^\infty du \frac{u^4}{\sqrt{u^2+\nu^2}}\frac{1}{\exp{[\sqrt{u^2 + \nu^2} - a]} + 1}\, , 	
\end{equation}
with $\nu=m_q/T$,  $\mu_q$ the quark chemical potential, $m_q$ the
bare quark mass, $T$  the temperature, and 
\begin{equation}
\label{presiong}
	P_g = \frac{8 T^4}{3 \pi^2}  \int_0^\infty  d\chi \chi^3 \frac{1}{\exp{(\chi + \frac{9 V_1}{8T} )} - 1}  \, .
\end{equation}

Considering $u$, $d$ and $s$ quarks, the total quark-gluon plasma pressure, $P_{qg}$, is given by
\begin{equation}
	P_{qg} = \sum_{q=u,d,s} (P_q+P_{\bar{q}}) + P_g + P_{vac} \, ,
\end{equation}
where, $P_{\bar{q}}$ is the anti-quarks contribution and $P_{vac}$ is
the vacuum pressure given by \citep{Simonov:2007a,Simonov:2007b,Nefediev:2009}
\begin{equation}
	P_{vac} = - \frac{(11 - 2/3 N_f)}{32} \frac{G_2}{2} \, , 
\end{equation}
where $N_f=3$ is the number of flavors.

Since Eq.(\ref{presionq}) coincides with the pressure for a free Fermi-Dirac quark gas in the limit $V_1=0$, we built a double power series expansion in terms of $m_q^2 /(u^2 T^2 + m_q^2)^2$ and $(\mu_q - V_1/2)/T$ to calculate  Eq.(\ref{presionq}) semi-analytically. The calculation was made by adapting the approximation developed in the work of \cite{Masperi:2004}, for a strange free quark gas at finite temperature. Details of the calculation are given in Appendix \ref{apendice1}. 

On the other hand, Eq.(\ref{presiong}) can be solved analytically:
\begin{equation}
  P_g = 16 \, \frac{T^4}{\pi^2}  % \rm{PolyLog}[4,e^{-\frac{9}{8}\frac{V_1}{T}}]
     \mathrm{Li}_4\left(e^{-\frac{9}{8}\frac{V_1}{T}}\right) 
        \, ,
\end{equation}
where $\rm{Li}_s[z]$ is the $Jonqui\grave{e}re's$ $function$.

The energy density $\epsilon_{qg}$ reads

\begin{equation}
\epsilon_{qg} = -P_{qg} + \sum_{q=u,d,s} \mu_q n_{qg} + T S_{qg},
\end{equation}
where $n_{qg}= \frac{\partial P_{qg}}{\partial \mu_q}$, is the quark
number density and $S_{qg} = \frac{\partial P_{qg}}{\partial T}$, is 
the entropy density.

The large distance $q\bar q$ potential, $V_1$, is given by
\begin{equation}
\label{ve1}
V_1 = \int_0^{1/T} d\tau(1-\tau T) \int_0^\infty d\chi \chi D_1^E(\sqrt{\chi^2 + \tau^2}) \, .	
\end{equation}

It can be shown \citep{Kuzmenko:2004} that since at large separations the string acts on quark with the force $\bar\sigma$, which is the string tension related with $\Lambda_{QCD}$, then $D_1^E$ in Eq.(\ref{ve1}) can be normalized according to $D_1^E(\xi) \sim \frac{\bar\sigma}{\pi \lambda^2}$
$\exp(-\mid\xi\mid/\lambda)$ being $\lambda \sim \unit[0.2]{fm}$ the correlation length of QCD vacuum in accordance with lattice results, and
$\bar\sigma \simeq \unit[0.18]{GeV^2}$.

Even though the phenomenological value of $V_1$ and $G_2$ are constrained by lattice calculations at zero baryonic density and finite temperature, the behavior of quark gluon-plasma at low temperatures and large chemical potential, as occurs in the core of NS, could be very different \citep{Burgio:2016}. On the other hand, the sum rules of the QCD have determined a numerical value for $G_2$, but it has a large uncertainty: $G_2 = \unit[(0.012 \pm 0.006)] {GeV}^4$ \citep{Shifman:1979b, Shifman:1979a}. 

As the static potential $V_1$ depends on the temperature, we have analyzed its variation in a range corresponding to the temperatures obtained for proto-hybrid stars. Solving Eq.(\ref{ve1}), $V_1$ can be rewritten as
\begin{equation}
\label{v1devel}
V_1 =\frac{\bar\sigma \lambda}{\pi} \left[ \; (3 \lambda T + 1) \; e^{-\frac{1}{\lambda T}}- 3 \lambda T + 2 \;\right] \,\,,
\end{equation}
in agreement with the result obtained by \cite{Bombaci:mnras} in Eq.(9), if one considers $V_1(0) =\frac{2\bar\sigma \lambda}{\pi} $ and $\hbar c =1$. At this point, is is worth noting that if we take $V_1(T_c) = 0.5 GeV$,  which is a value independent of the number of quark flavors (see \cite{Simonov:2007a} and \cite{Simonov:2007b}) we obtain a critical temperature $T_c \sim 270 MeV$,  the critical temperature of a hot pure gluon plasma without quarks (see \cite{Simonov:1999} and \cite{glueballs:2016} for details). However, for $T_c \sim 150 MeV$, corresponding to the chiral transition, the approximately value of $V_1(0)$ should be $V_1(0) \sim \unit[0.65] {GeV}$ according to Eq. (\ref{v1devel}). Additionally, the potential $V_1$  and the gluon condensate $G_2$ are related for the deconfinement transition at zero chemical potential at the critical temperature, $T_c$, and can be calculated through  \citep{Bombaci:mnras}
 \begin{equation}
T_c=\frac{a_0}{2} G_2^{1/4} \left( 1+\sqrt{1+\frac{V_1(T_c)}{2 a_0 G_2^{1/4}}} \right),
 \end{equation}
where, $a_0 \sim 0.44$ for three flavors. 

On the other hand (except at $T_c$), $G_2$ is essentially independent on the temperature \citep{Burgio:2016}. It is important to note at this point, the analysis about the dependence of  $G_2$ and $V_1$ on the temperature is at vanishing chemical potential. There are no simulations of lattice QCD at high chemical potential, and at finite chemical potential standard lattice methods can not be used because the sign problem of QCD at non-zero baryon density \citep{Fodor:2002}. Additionally, for the range of temperatures obtained in this work  in the framework of proto-NS, $\unit[0] {MeV} < T < \unit[60] {MeV}$, we found that the variation of $V_1$ is only $0.2\%$ and it does not have any impact on the results.  A detailed study of $V_1$ related to the lattice QCD and the measured NS masses has been carried out by \cite{Bombaci:mnras} and \cite{Plumari:2013}.

Thereby, attempting to impose any restrictions to $V_1$ and/or $G_2$ in a regime of low temperatures and high densities  we will treat them as free parameters, independent of the temperature and/or the baryonic chemical potential. We have chosen three representative sets of parameters in the $V_1-G_2$ plane: Set 1= ($V_1 = \unit[0.02] {GeV}$, $G_2 = \unit[0.016] {GeV}^4$), Set 2 = ($V_1 = \unit[0.03] {GeV}$, $G_2 = \unit[0.010] {GeV}^4$) and Set 3 =  ($V_1 = \unit[0.10] {GeV}$, $G_2 = \unit[0.007] {GeV}^4$).  The choice of these sets also ensure a minimum maximum mass value of $\sim 2 M_{\odot}$ in the M-R diagram for cold NS.

%----------------------------------------------------------------------------------------

\subsection{Stability of quark matter}

Since \cite{Bodmer} and \cite{Witten} opened up the field with the hypothesis of strange quark matter, many authors have studied and discussed the possibility that the inner core of compact objects is composed of such form of matter (\cite{Glendenning:1996, Smith:2010} and references therein).

The strange quark matter hypothesis suggest that strange quark matter would be the true ground state of strong interactions, that is, its energy per baryon should be less than that of the isotope $^{56}\mathrm{Fe}$ ($\sim \unit[930]{MeV}$), the most bound nucleus. This hypothesis can be explained easily through the MIT Bag model if one consider non-interacting massless quarks. But this is an oversimplified treatment of the issue, because the effect of a finite strange quark mass, as well as a non-zero strong coupling constant, constraints the value of the Bag constant in the model to get stable quark matter \citep{Weber2005}. In addition, studies of dense quark matter within more realistic models, like the Nambu-Jona-Lasinio (NJL) model, do not support the idea of absolutely stable strange quark matter \citep{Buballa:2005}. The main reason is the dynamical mass of the strange quark.

If massless $u$, $d$ and $s$ quarks at zero temperature and pressure are considered in the MIT Bag model, strange quark matter would be more stable than non-strange quark matter because the conversion of about one third of $u$ and $d$ quarks into $s$ quarks via weak interactions, decreases the energy of the system, due that baryon number is shared among three Fermi seas instead of two. But if the matter is described considering dynamical quark masses, like in the NJL model, the strange quark mass remains still large at densities where chiral symmetry is almost restored in the non-strange sector, and then non-strange quark matter tends to be energetically more favored than strange quark matter. The conversion of a $d$ quark into a $s$ quark costs too much energy.

It has been suggested that there would be other phases of strange quark matter, color superconducting phases like the color-flavor locked (CFL), more favorable to the stability condition due the emergence of diquark condensates \citep{Lugones:2002}. However, although the formation of diquark condensates reduces the energy per baryon, it is still larger than the $^{56}\mathrm{Fe}$  if dynamical quark masses are considered \citep{Buballa:2005}.

Recent studies about the strange quark matter hypothesis in the framework of the FCM \citep {Pereira:2013}, show that for some choices of the parameters $G_2$ and $V_1$, the energy per baryon can be lower than the energy per nucleon of the $^{56}\mathrm{Fe}$. According to \cite{Pereira:2013}, the stability window strongly depends on the values of the FCM parameters: as $V_1$ increases, from $V_1 =\unit[0]{ GeV}$ to $V_1 = \unit[0.5] {GeV}$, $G_2$ must decrease in order to keep the energy per baryon of the quark matter less than $\unit[930]{MeV}$.

\begin{figure}[tb]
	\centering
	\includegraphics[width=0.95\columnwidth]{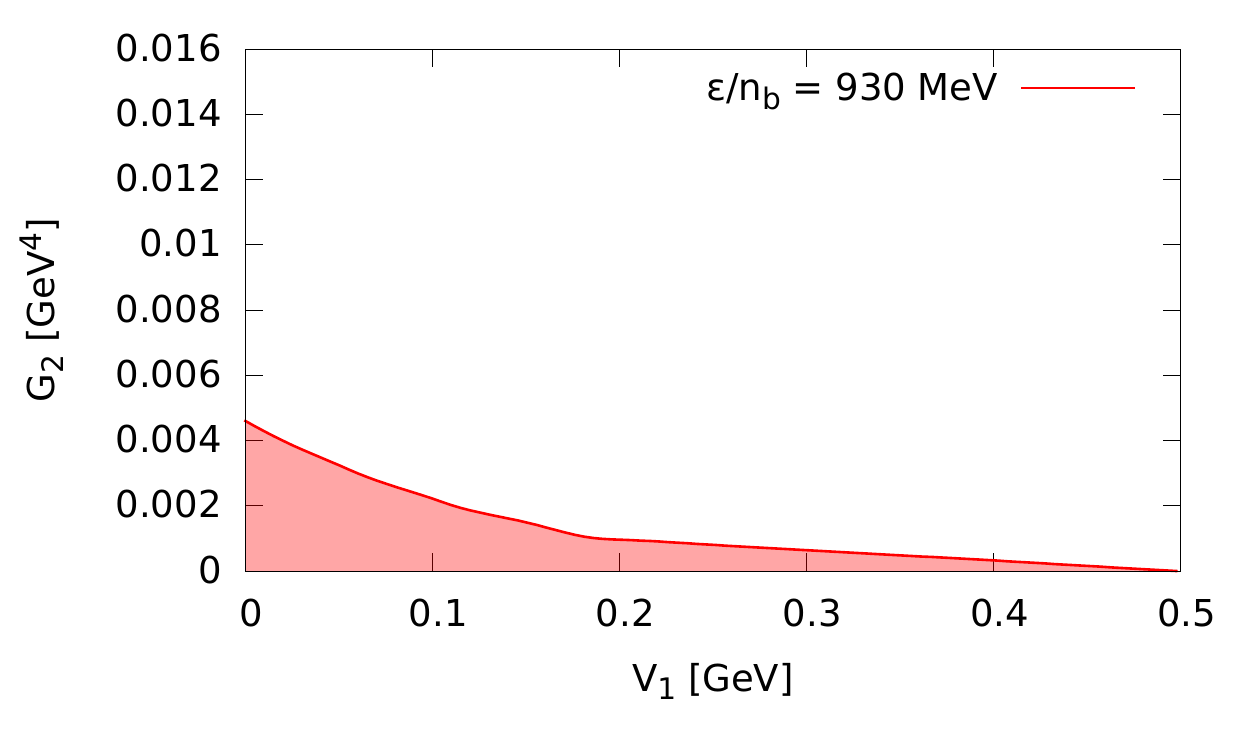}
	\caption{Stability window in
          the $G_2-V_1$ plane. The colored area under the $\epsilon/n_b = 930$ MeV curve
          indicates the stability region of the strange quark matter
          ($m_u = \unit[2.3] {MeV}$, $m_d = \unit[4.8] {MeV}$,
          $m_s = \unit[95] {MeV}$) at zero temperature and
          pressure. The solid line indicates where the energy per
          baryon of the quark matter matches with the energy per
          nucleon in the $^{56}\mathrm{Fe}$ most bound nucleus. Considering the QCD sum
          rules, the estimated value of
          $G_2 = \unit[(0.012 \pm 0.006)] {GeV}^4$ is out of the
          stability region.}
	\label{stability}
\end{figure}

We have also analyzed the stability of pure quark matter under our semi-analytical approach. Taking the quark masses $m_u = \unit[2.3] {MeV}$,
$m_d = \unit[4.8] {MeV}$ and $m_s = \unit[95] {MeV}$, we have calculated the energy per baryon at zero temperature and pressure. The results are shown in Fig. \ref{stability}. We obtain a narrow stability window which is very sensitive to the FCM parameters, in agreement with the results of \cite{Pereira:2013}.

From Fig. \ref{stability} it is evident that the values of $G_2$ associated with the stability of strange quark matter are out of
the range of values estimated through the QCD sum rules, $G_2 = \unit[(0.012 \pm 0.006)] {GeV}^4$. However, we found that it is possible to obtain a strange star (SS) with $M \geq 2 M_{\odot}$ by combining $V_1$ and $G_2$ values inside the stability region. For example, taking $V_1 = \unit[0.1] {GeV}$ and $G_2 = \unit[0.002] {GeV}^4$, we obtain a set of cold stable SS with $M_{max} = 2.01 M_{\odot}$.

In this work we consider a metastable quark matter phase, taking $G_2 = \unit[(0.012 \pm 0.006)] {GeV}^4$ according with the QCD sum rules, but out of values determining the stability window.

%==========================================================================================================
\section{General conditions for the hybrid EoS and phase transition}
\label{phase}
 
Lattice gauge simulations at finite temperature in QCD quark matter \citep{Yasutake:2013} suggests a range of the surface tension $\sim \unit[10 -100] {MeV\,fm^{-2}}$ between the interface hadron-quark. If the surface tension is bigger than a critical value, which is estimated to be $ \sim \unit[40] {MeV\,fm^{-2}}$ then a sharp phase transition, between hadronic and quark phases, is favored via Maxwell construction \citep{Endo:2006, Yasutake:2014}.

Due the uncertainty in the value of the surface tension we assume an abrupt phase transition between hadronic and quark matter. Thus, in
order to assemble the hybrid EoS, we work under the Maxwell construction, in which the crossing of both hadronic and quark EoS curves in the $\mu-P$ plane defines (or not) the phase transition.

Baryonic and electric charges are conserved locally. For the inner core of the HS, such conditions are given by
\begin{eqnarray}
3n_B-n_u-n_d-n_s = 0 \, ,\, \, \\ \nonumber
  2n_u - n_d - n_s - 3 (n_e +n_{\mu}) = 0 \, ,
\end{eqnarray}
where $n_u$, $n_d$ and $n_s$ are the quark number densities, $n_e$ and $n_{\mu}$ are electron and muon densities, respectively, and $n_B$ is the baryonic number density. 

Beta-equilibrium condition determines the following relationships between the chemical potentials of the particles involved:
\begin{equation}
  \mu_u = 1/3 \mu_b - 2/3 \mu_e + 2/3 \mu_\nu,
\end{equation}
\begin{equation}
  \mu_d = \mu_s = 1/3 \mu_b + 1/3 \mu_e - 1/3 \mu_\nu, \nonumber
\end{equation}
where $\mu_{i=u,d,s}$ are the quark chemical potentials and the subscripts $b$, $e$ and $\nu$ indicates baryon, electron and neutrino chemical potentials, respectively.

The outer core EoS is composed by nucleonic degrees of freedom (neutrons and protons), electrons, muons and neutrinos in beta-equilibrium with local charge conservation \citep{Shen:2011}.

By setting equal pressure and baryonic chemical potential during the phase transition in the NeStOR code, we establish the conditions to determine the thermodynamic properties of HHS and CHS cores.

\begin{figure}[tb]
	\centering
	\includegraphics[width=0.85\columnwidth]{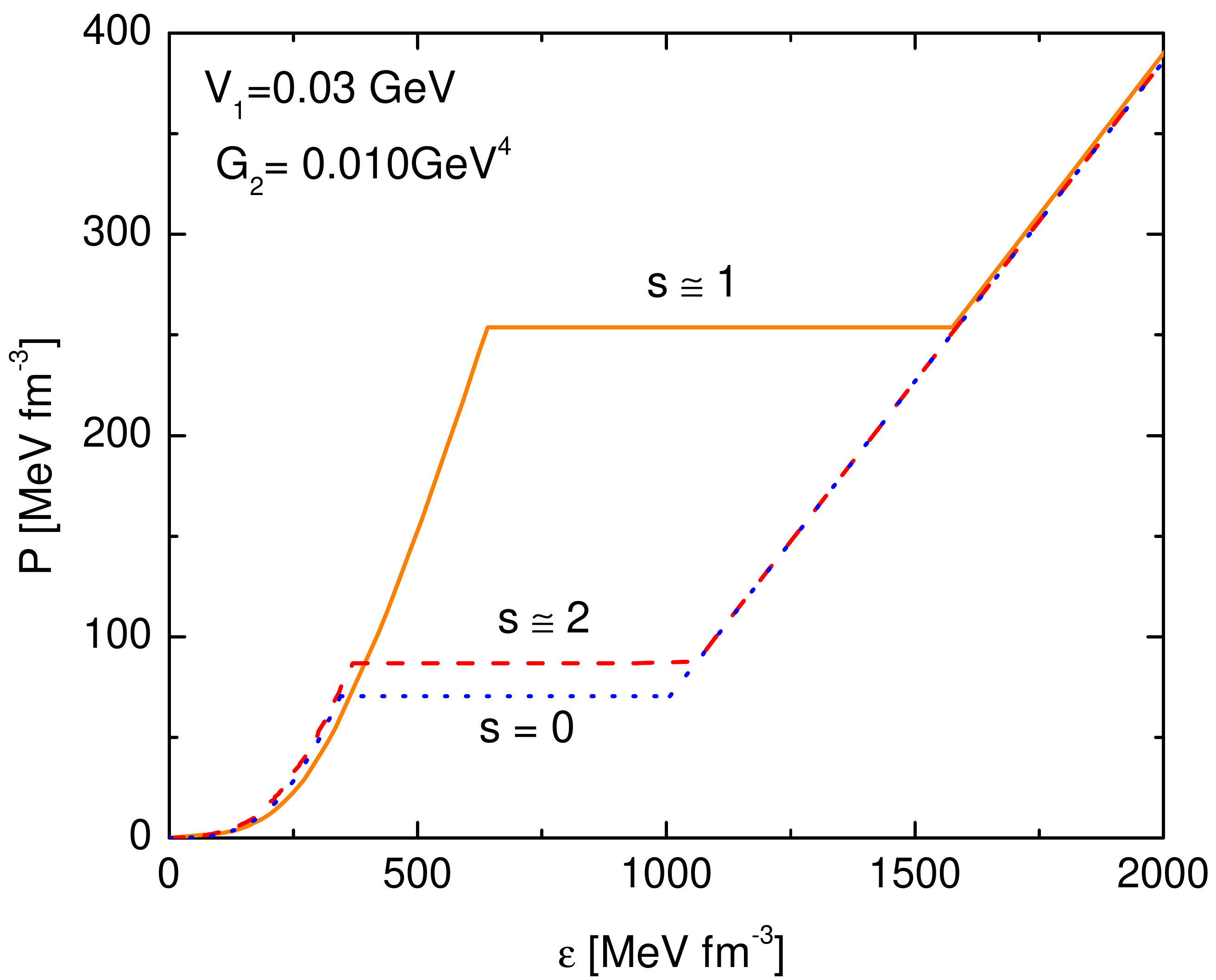}
	\caption{Hybrid EoS for a particular pair of the
          FCM parameters, $V_1 = \unit[0.03] {GeV}$ and
          $G_2 = \unit[0.010] {GeV}^4$ (Set 2), for the three stages of
          constant entropy per baryon. 
        }
	\label{eos3s}%
\end{figure}

The EoS for the different stages of thermal evolution for the Set 2 are shown in Fig. \ref{eos3s}. As the star cools, the EoS becomes stiff. The baryonic densities at the transition points are $n_B \sim \unit[0.53] {\, fm^{-3}} $ for $s \simeq 1$, $n_B \sim \unit[0.34] {\, fm^{-3}} $ for $s \simeq 2$, and $n_B \sim \unit[0.33] {\, fm^{-3}} $ for $s = 0$. In the case of $s \simeq 1$ although neutrinos soften the EoS, their contribution shifts the transition to higher densities. The transition pressure are $P_t = \unit[253.79] {MeV\, fm^{-3}}$, for $s \simeq 1$ and $P_t = \unit[86.87] {MeV\, fm^{-3}}$ and $P_t = \unit[70.41] {MeV\, fm^{-3}}$ for $s \simeq 2$ and $s = 0$ respectively. The behavior of the EoS for Set 3 is similar to that of Set 1.

%----------------------------------------------------------------------------------------

For the outer core of the HS we have used the hadronic EoS constructed by \cite{Shen:2010a}. This EoS is built using the 
relativistic mean field (RMF) model with a density dependent coupling that is a slightly modified form of the original NL3 interaction (see \cite{Shen:2010a} for details). The outer core is composed of neutrons, protons and electrons and we have included neutrinos in the NeStOR code. The complete EoS for astrophysical simulations in proto-NS is detailed in \cite{Shen:2011}.

For nuclear matter at sub-nuclear densities we have considered the work of \cite{Shen:2010b}, where a Virial expansion for a non-ideal gas composed of neutrons, protons, alpha particles, and nuclei is used to build the low density EoS at finite temperature. The minimum energy densities obtained through the NeStOR code were $\epsilon \sim 2 \times \unit[10^{10}] {g/cm^3}$ for $s \sim 1$, $\epsilon \sim 4 \times \unit[10^{7}] {g/cm^3}$ for $s \sim 2$ and $\epsilon \sim 2 \times \unit[10^{7}] {g/cm^3}$ for  $s = 0$. For lower density values, we have considered the Baym–Pethick–Sutherland (BPS) crustal model \citep{Baym:1971a}.

The leptonic contribution is provided by electrons, electron-neutrinos and muons. For electrons and muons we have considered the same expressions used
for quark matter, under $s = 0$ and $s \simeq 2$ scenarios, making the correspondent correction in the degeneracy factor and setting $V_1=0$, which mimics exactly a Fermi-Dirac gas at finite temperature. Under  $s \simeq1$ scenario, we had some numerical instabilities due the inclusion of electron-neutrinos. To avoid this, we use a massless finite-temperature Sommerfeld expansion for the electrons \citep{Lugones:2009},
\begin{equation}
\label{somer}
 	P_e= \frac{\mu_e ^4+\frac{7 \pi ^4 T^4}{15}+2 \pi ^2 \mu_e ^2 T^2}{12 \pi ^2} \, ,
\end{equation}
which simplifies the numerical calculation without loss of
generality. For the electron-neutrinos we have used the same
expansion taking into account the appropriate correction in the degeneracy factor.

\begin{figure}[tb]
	\centering
	\includegraphics[width=0.90\columnwidth]{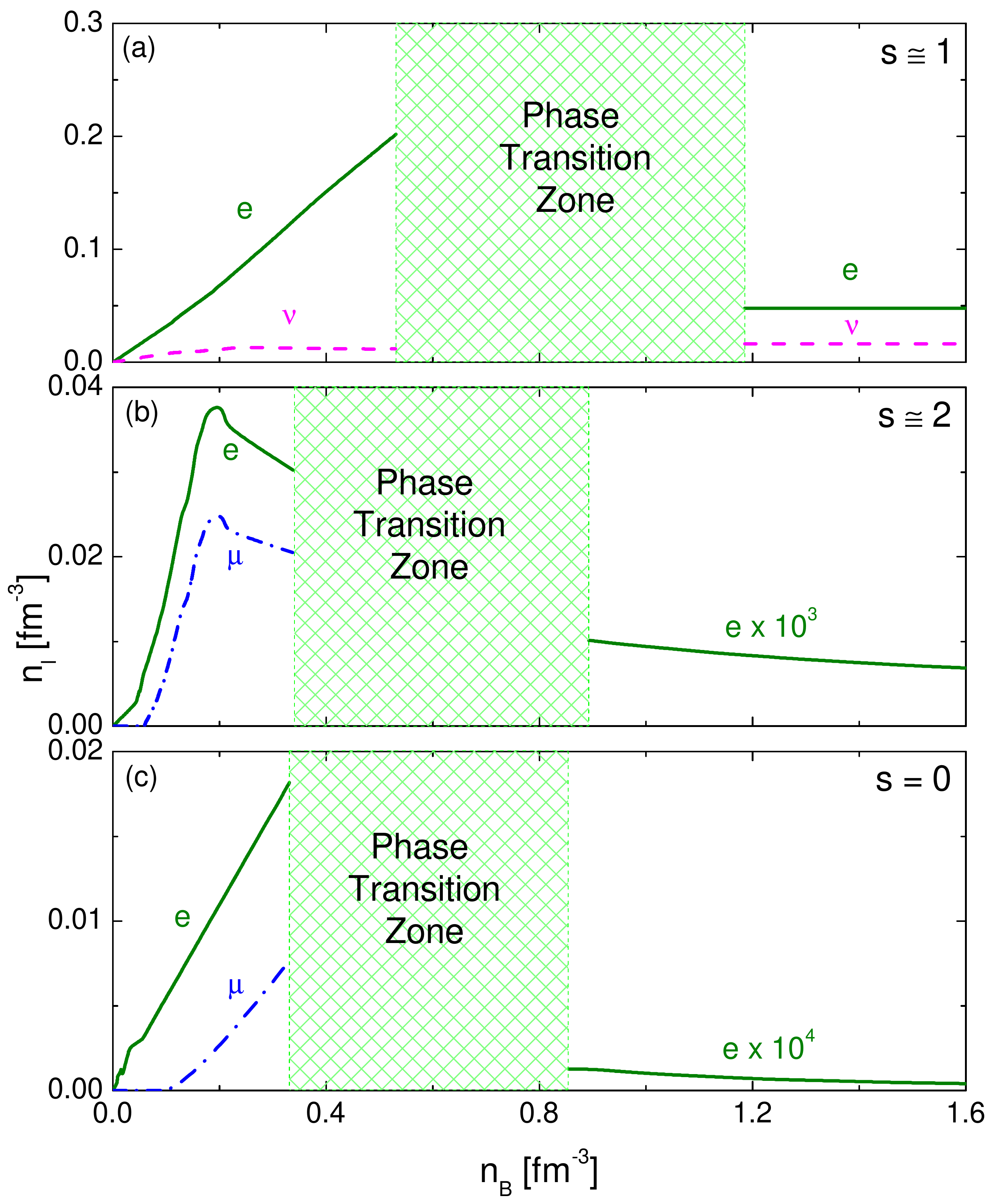}
	\caption{Lepton densities $n_{l}$ as a function of baryonic
          density, $n_B$, considering the Set 2  for the three fixed entropies: (a)
          $s \simeq 1$, with lepton abundances $Y_{L_e} = 0.4$ and
          $Y_{L_\mu}= 0$, (b) $s \simeq 2$, and (c) $s =0$. Hatched areas indicate the phase transition zone. In panels
          (b) and (c) $Y_{\nu} = 0$, and muons are present.}
	\label{leptons}%
\end{figure}

Figure \ref{leptons} shows the lepton densities for the different fixed entropy stages considered, taking the Set 2 as an example. In Fig. \ref{leptons} (a), it can be seen that the lepton density reaches values relatively high ($\sim$ 40 $\%$ $n_B$) before the zone of the phase transition due there is also a high proton fraction in the hadronic phase. The contribution of neutrinos to the pressure shifts the phase transition to quark matter at higher densities, because the inclusion of more degrees of freedom in the system soften the EoS and therefore is more difficult to reach the pressure of quark matter at high densities. We should remark that the high abundance of neutrinos and their contribution to the pressure is the result of the condition, $Y_{L_e} = 0.4$. At this stage we obtain that neutrinos density, $n_{\nu}$, goes from zero to $\sim \unit[0.02] {fm^{-3}}$, almost the same maximum density reached by the muons in the following stage. Electrons and neutrinos are also present in the quark phase. In Fig. \ref{leptons} (b), the leptonic fraction decreases in the left side of the phase transition zone, and after the peak in the hadronic region the proton fraction drop off and matter gets neutron rich. There is very little amount of electrons in the quark phase. The entropy zero stage at the left side of Fig. \ref{leptons} (c) shows a decrease in the density of electrons and muons for the hadronic phase. For the quark phase, muons do not contribute and the density of electrons is negligible. For Set 1 and Set 3 the analysis of the lepton densities is similar.

%==========================================================================================================

\section{Snapshots of thermal evolution for proto-hybrid stars}
\label{entop}

Until the proto-NS reaches the thermal equilibrium, it undergoes a process of thermal evolution. The evolutionary scenario is based on a dynamical calculation with schematic EoS (see \cite{Prakash:1997} and references therein). The different stages of such process are studied by considering three sequential snapshots of constant entropy per baryon. According the thermodynamic properties, each stage can be
characterized as follows \citep{Burrows:1986, Steiner:2000, Shao:2011}. First, when the NS is born, neutrinos created in beta reactions are trapped because their mean free path is small compared with the star radius and there is leptonic abundance conservation in the star matter.  No muons are present because trapped neutrinos inhibit their existence. Therefore, this stage is characterized by the conditions
 	\begin{equation}
 		Y_{L_e} \equiv \frac{n_e + n_{\nu_e}}{n_B} = 0.4 \ ; \
                Y_{L_\mu} \equiv \frac{n_\mu + n_{\nu_\mu}}{n_B} = 0 \ , 
 	\end{equation}
and entropy per baryon, $s \approx 1$.
 	
After $\sim 10-20$ seconds, neutrino diffusion deleptonized the star matter and the presence of muons becomes relevant. Thus, regarding the lepton abundances,
 	\begin{equation}
 		Y_{\nu_e} \equiv \frac{n_{\nu_e}}{n_B} = 0 \ ; \
                Y_{\nu_\mu} \equiv \frac{n_{\nu_\mu}}{n_B} = 0 \ . 
 	\end{equation}
Neutrino diffusion yield most of its energy to the proto-NS, heating it. At this stage, the entropy per baryon reaches $s \approx 2$.
 	
Finally, after a few minutes, the star cools down and it can be characterized by an entropy per baryon, $s = 0$. At this stage, there is also absence of neutrinos, with
 	\begin{equation}
 		Y_{\nu_e} \equiv \frac{n_{\nu_e}}{n_B} = 0 \ ; \
                Y_{\nu_\mu} \equiv \frac{n_{\nu_\mu}}{n_B} = 0 \ . 
 	\end{equation}

\begin{figure}[tb]
  \centering
  \includegraphics[width=0.85\columnwidth]{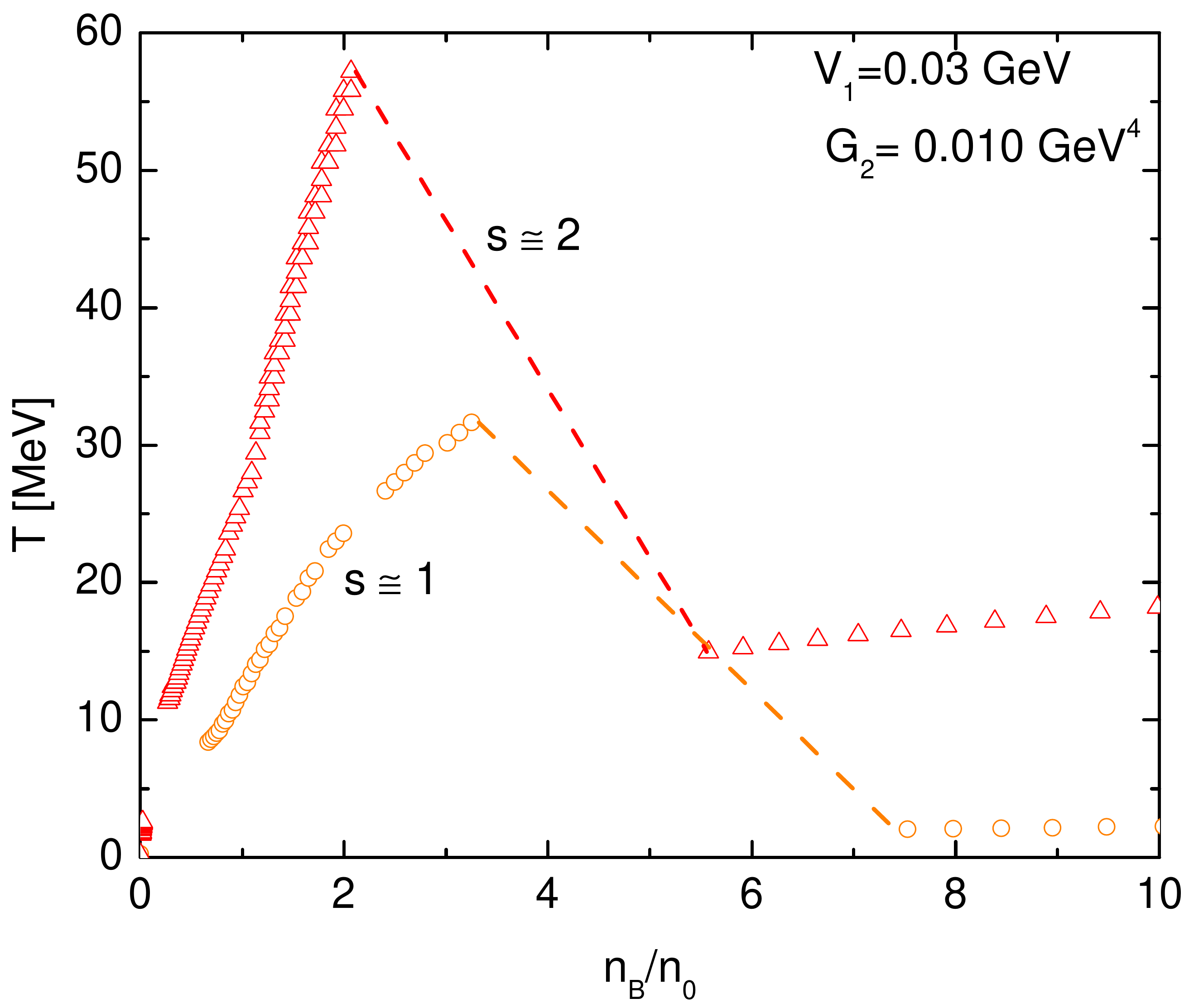}
  \caption{Snapshots of the temperature as a function of baryonic
    density in units of the saturation nuclear density for the Set 2.} 
  \label{temps}%
\end{figure}

We show the evolution of the temperature as a function of the baryonic density in Fig. \ref{temps}, Set 2, for entropies $s \neq 0$ . When $s \simeq 1$ the maximum temperature reached is $T = \unit[31.66] {MeV}$, whereas if $s \simeq 2$ the maximum temperature corresponds to $T = \unit[57.19] {MeV}$. The evolution of the temperatures for Set 3 is similar to that of Set 1.

Thereby, we can characterize three different stages of the NS thermal evolution, in a simplified way, through an isentropic state. This approach allow us to study the structure and composition of NS modeled as HS during their thermal evolution on their first minutes of
existence.

%==========================================================================================================

\section{Structure of hybrid stars}
\label{hybrid}

Once the hybrid EoS is obtained, we have the necessary input to solve the Tolmann-Oppenheimer-Volkoff (TOV) equations, which are the relativistic
structure equations of hydrostatic equilibrium and mass conservation for a spherically symmetric configuration given by
\begin{equation}
\label{tov1}
\frac{dP}{dr} = -\frac{G m(r) \epsilon(r)}{r^2} \frac{ [1 + P(r) / \epsilon(r) ] [1 + 4 \pi r^3 P(r) / m(r)]}{1-2 G m(r)/r} \,,
\end{equation}
\begin{equation}
\label{tov2}
\frac{dm}{dr} = 4 \pi r^{2} \, \epsilon(r) \,,	
\end{equation}
where $r$ is the radius, $P(r)$ and $\epsilon(r)$ are the pressure and energy density at a radius $r$, $m(r)$ is the mass bounded by the radius
$r$ and $G$ is the universal gravitational constant.

Equations (\ref{tov1}) and (\ref{tov2}) are integrated solving the HS structure, giving for each set of parameters a family of stars with
different initial conditions for the central energy density. Thus, the EoS determines the mass-radius relationship for each group of stars.

\begin{figure}[tb]
	\centering
	\includegraphics[width=0.95\columnwidth]{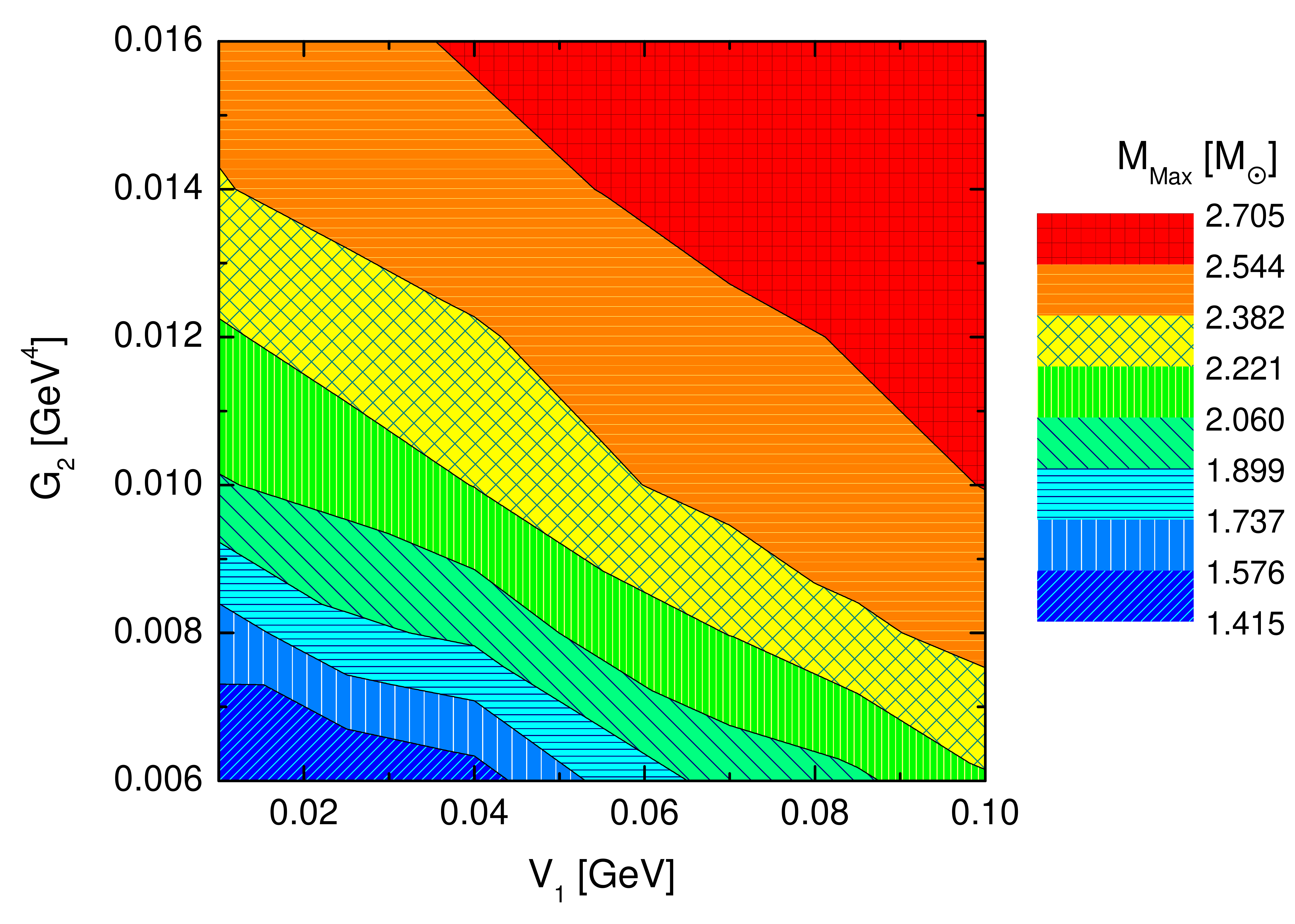}
	\caption{Maximum mass of each family of stars as a function of $V_1$
          and $G_2$ at $s = 0$. Inside each contour area, the maximum
          mass increases monotonically as the values $V_1$ and $G_2$ increase.}
      \label{contour}%
\end{figure}

To solve Eqs. (\ref{tov1}) and (\ref{tov2}), and following the works of \cite{Logoteta:2013,Burgio:2016,Plumari:2013}, we have chosen several combinations of $V_1$ ($\unit[0.01] {GeV} < V_1 < \unit[0.10] {GeV}$) and $G_2 = \unit[(0.012 \pm 0.006)] {GeV^4}$. By combining these range of values for the FCM parameters as inputs in the NeStOR code, we obtain several data and tables as outputs of the program. The main microscopic-thermodynamic results
are the hybrid EoS, particle abundances-density and temperature-density relationships. The macroscopic-astrophysics results are the mass-radius, the mass-central energy density and the gravitational mass-baryonic mass relationships. In the $V_1-G_2$ plane, we also construct contours of the maximum mass as is shown in Fig. \ref{contour}.

\begin{figure}[tb]
	\centering
	\includegraphics[width=0.85\columnwidth]{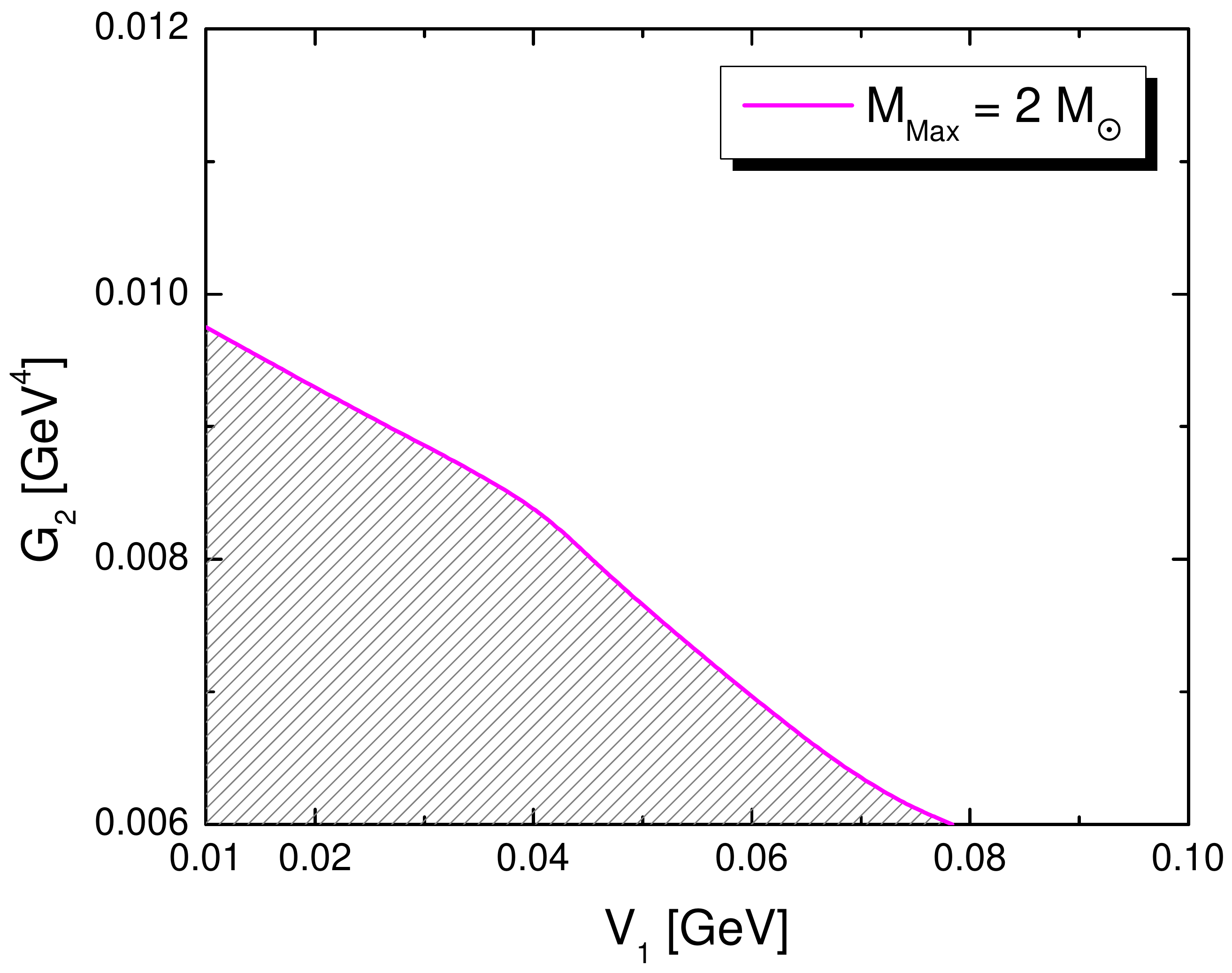}
	\caption{Contour line for $M_{max} = 2 M_\odot$ as a function of
          $V_1$ and $G_2$ at $s = 0$.  Shaded region below the curve indicates stars with masses $< 2 M_\odot$. }
	\label{line}%
\end{figure}

For the stage corresponding to zero entropy, a $V_1-G_2$ plane was constructed to determine the families of HS in which it is possible to obtain a minimum maximum mass star of $2 M_\odot$. In Fig. \ref{contour}, we show the variation of the maximum mass as a function of the FCM parameters $V_1$ and $G_2$. We obtain contours from $M_{max} \approx 1.5 M_\odot$ to $M_{max} \approx 2.7 M_\odot$ depending on the combination of the parameters. In Fig. \ref{line}, we show in detail the $M_{max} = 2.0 M_\odot$ contour.

Table \ref{tabla} shows the masses and radii computed for different combinations of the FCM parameters. We have used the upper and lower bounds for both maximum masses and radii obtained, corresponding to the four corners of Fig. \ref{contour}. We also have chosen these parameter values to determine the shaded area delimited by the solid curves in the phase diagram of Fig. \ref{phase_diag}.

\begin{table}[tb]
\centering
\caption{Maximum mass, $M_{max}$, and radius, $R_{M_{max}}$, for the extreme values of the FCM parameters, $V_1$ and $G_2$, at different fixed  entropy per baryon.}
\label{tabla}
\begin{tabular}{|l|l|l|l|l|}
\hline
$s \, (k_B)$           & $V_1 ($GeV$)$          & $G_2 ($GeV$^4)$          & $M_{max} (M_{\odot})$ & $R_{M_{max}} (km)$ \\ \hline
\multirow{8}{*}{1} & \multirow{4}{*}{$0.01$}  & \multirow{2}{*}{$0.006$} & \multirow{2}{*}{2.40}     & \multirow{2}{*}{14.3}  \\
                   &                        &                          &                       &                    \\ \cline{3-5} 
                   &                        & \multirow{2}{*}{$0.016$} & \multirow{2}{*}{2.68}     & \multirow{2}{*}{13.6}  \\
                   &                        &                          &                       &                    \\ \cline{2-5} 
                   & \multirow{4}{*}{$0.10$} & \multirow{2}{*}{$0.006$} & \multirow{2}{*}{2.69}     & \multirow{2}{*}{13.5}  \\
                   &                        &                          &                       &                    \\ \cline{3-5} 
                   &                        & \multirow{2}{*}{$0.016$} & \multirow{2}{*}{2.72}     & \multirow{2}{*}{13.2}  \\
                   &                        &                          &                       &                    \\ \hline
\multirow{8}{*}{2} & \multirow{4}{*}{$0.01$}  & \multirow{2}{*}{$0.006$} & \multirow{2}{*}{1.44}     & \multirow{2}{*}{10.0}  \\
                   &                        &                          &                       &                    \\ \cline{3-5} 
                   &                        & \multirow{2}{*}{$0.016$} & \multirow{2}{*}{2.63}     & \multirow{2}{*}{15.5}  \\
                   &                        &                          &                       &                    \\ \cline{2-5} 
                   & \multirow{4}{*}{$0.10$} & \multirow{2}{*}{$0.006$} & \multirow{2}{*}{2.49}     & \multirow{2}{*}{15.8}  \\
                   &                        &                          &                       &                    \\ \cline{3-5} 
                   &                        & \multirow{2}{*}{$0.016$} & \multirow{2}{*}{2.80}     & \multirow{2}{*}{15.0}  \\
                   &                        &                          &                       &                    \\ \hline
\multirow{8}{*}{0} & \multirow{4}{*}{$0.01$}  & \multirow{2}{*}{$0.006$} & \multirow{2}{*}{1.42}     & \multirow{2}{*}{9.3}  \\
                   &                        &                          &                       &                    \\ \cline{3-5} 
                   &                        & \multirow{2}{*}{$0.016$} & \multirow{2}{*}{2.43}     & \multirow{2}{*}{15.1}  \\
                   &                        &                          &                       &                    \\ \cline{2-5} 
                   & \multirow{4}{*}{$0.10$} & \multirow{2}{*}{$0.006$} & \multirow{2}{*}{2.20}     & \multirow{2}{*}{15.2}  \\
                   &                        &                          &                       &                    \\ \cline{3-5} 
                   &                        & \multirow{2}{*}{$0.016$} & \multirow{2}{*}{2.70}     & \multirow{2}{*}{14.7}  \\
                   &                        &                          &                       &                    \\ \hline
\end{tabular}
\end{table}

\begin{figure}[tb]
	\centering
	\includegraphics[width=0.95\columnwidth]{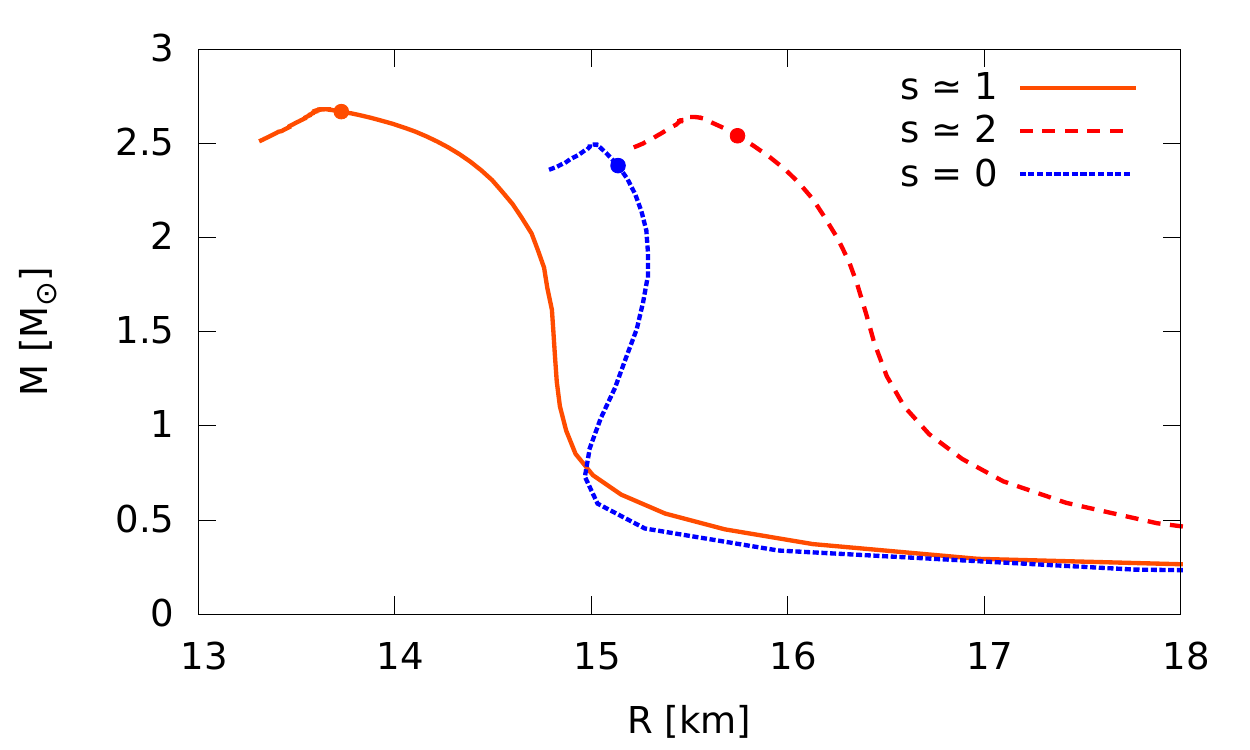}
	\caption{M-R relationship for Set 1 ($V_1 = \unit[0.02] {GeV}$, $G_2 = \unit[0.016] {GeV}^4$). In each constant-entropy curve, the rounded dot
          indicates the appearance of the quark matter core.}
	\label{mraio1}%
\end{figure}

\begin{figure}[tb]
  \centering
  \includegraphics[width=0.95\columnwidth]{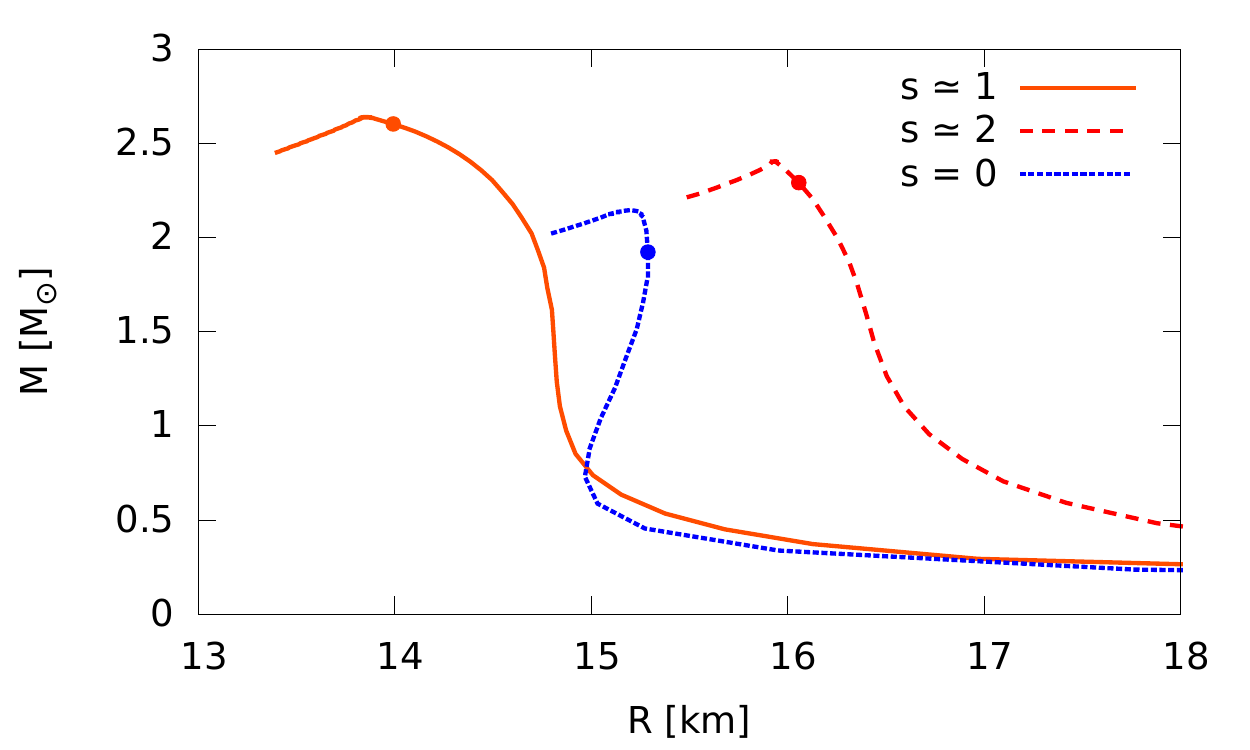}
  \caption{M-R relationship for Set 2 ($V_1 = \unit[0.03] {GeV}$, $G_2 = \unit[0.010] {GeV}^4$). In each constant-entropy curve, the rounded dot
          indicates the appearance of the quark matter core.}
  \label{mraio2}%
\end{figure}

\begin{figure}[tb]
  \centering
  \includegraphics[width=0.95\columnwidth]{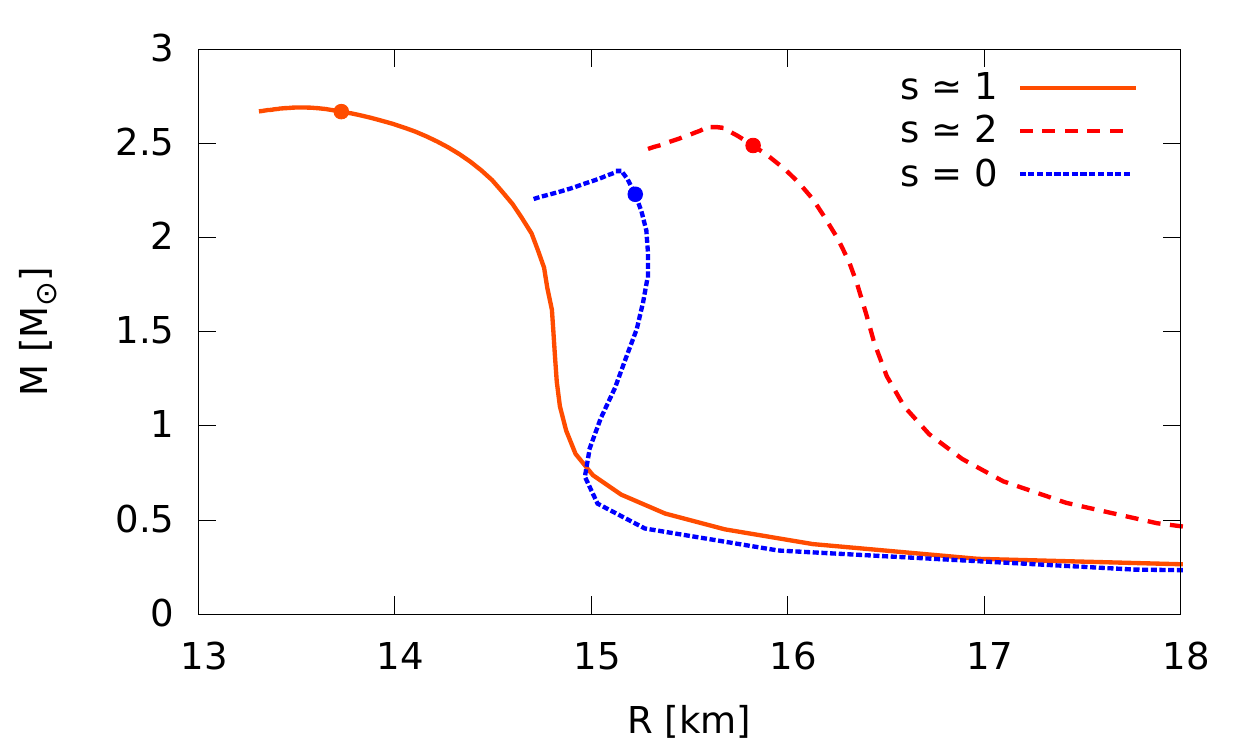}
  \caption{M-R relationship for Set 3 ($V_1 = \unit[0.10] {GeV}$, $G_2 = \unit[0.007] {GeV}^4$). In each constant-entropy curve, the rounded dot
          indicates the appearance of the quark matter core.}
  \label{mraio3}%
\end{figure}

As was mentioned before, we have considered different combinations of the parameters in the plane  $V_1-G_2$ in order to fulfill the $M_{max} \gtrsim 2 M_\odot$ condition. Besides this requirement, we have chosen three particular sets of parameters (Set 1, Set 2 and Set 3) so as to cover an extensive and representative region of different combinations in the $V_1-G_2$ plane. The mass-radius relationship for these sets and for $s \simeq 1$, $s \simeq 2$ and $s = 0$ are shown in Figs. \ref{mraio1}, \ref{mraio2} and \ref{mraio3}. It can be seen there are stable branches for the stellar structure up to a maximum value, $M_{max}$, from which unstable configurations are obtained as radii become smaller. Also, for each mass-radius relationship, there is a particular mass value, $M_{hyb}$, which indicates the phase transition between hadronic and quark matter and the appearance of HS. Table \ref{tabla2} shows the values of $M_{hyb}$ and $M_{max}$ obtained for the three sets.

\begin{table}[tb]
\centering
\caption{Maximum mass, $M_{max}$, and mass associated to the appearance of the quark matter core in the M-R relationship, $M_{hyb}$, for the three sets of the FCM parameters, $V_1$ and $G_2$, at different fixed entropy per baryon. Set 1 = ($V_1 = \unit[0.02] {GeV}$, $G_2 = \unit[0.016] {GeV}^4$), Set 2 = ($V_1 = \unit[0.03] {GeV}$, $G_2 = \unit[0.010] {GeV}^4$) and Set 3 =  ($V_1 = \unit[0.10] {GeV}$, $G_2 = \unit[0.007] {GeV}^4$).}
\label{tabla2}
\begin{tabular}{|l|l|l|l|}
\hline
                       & $s \, (k_B)$ & $M_{max} (M_{\odot})$ & $M_{hyb} (M_{\odot})$ \\ \hline
\multirow{3}{*}{Set 1}  & \multirow{2}{*}{$1$}         & \multirow{2}{*}{$2.68$}                  & \multirow{2}{*}{$2.67$}                  \\ 
                       &           &                       &                       \\ \cline{2-4} 
                       & \multirow{2}{*}{$2$}         & \multirow{2}{*}{$2.64$}                  & \multirow{2}{*}{$2.54$}                  \\
                       &           &                       &                       \\ \cline{2-4} 
                       & \multirow{2}{*}{$0$}         & \multirow{2}{*}{$2.49$}                  & \multirow{2}{*}{$2.38$}                  \\ 
                       &           &                       &                       \\ \hline
\multirow{3}{*}{Set 2} & \multirow{2}{*}{$1$}         & \multirow{2}{*}{$2.64$}                  & \multirow{2}{*}{$2.60$}                  \\ 
                       &           &                       &                       \\ 
                       &           &                       &                       \\ \cline{2-4} 
                       & \multirow{2}{*}{$2$}         & \multirow{2}{*}{$2.40$}                  & \multirow{2}{*}{$2.29$}                  \\ 
                       &           &                       &                       \\ \cline{2-4} 
                       & \multirow{2}{*}{$0$}         & \multirow{2}{*}{$2.14$}                  & \multirow{2}{*}{$1.92$}                  \\ 
                       &           &                       &                       \\ \hline
\multirow{3}{*}{Set 3} & \multirow{2}{*}{$1$}         & \multirow{2}{*}{$2.69$}                  & \multirow{2}{*}{$2.67$}                  \\ 
                       &           &                       &                       \\ \cline{2-4} 
                       & \multirow{2}{*}{$2$}         & \multirow{2}{*}{$2.59$}                  & \multirow{2}{*}{$2.49$}                  \\ 
                       &           &                       &                       \\ \cline{2-4} 
                       & \multirow{2}{*}{$0$}         & \multirow{2}{*}{$2.35$}                  & \multirow{2}{*}{$2.23$}                  \\ 
                       &           &                       &                       \\ \hline
\end{tabular}
\end{table}

We have computed the three stages of constant entropy for proto-HS through a simplified evolutionary process, from an initial state in which $s \simeq 1$ to a final state with $s = 0$. However, to make a suitable analysis of the proto-HS evolution, it is more appropriate to study the evolutionary process in the gravitational mass - baryonic mass plane, along a constant baryonic mass path.

Until now, we have used the term ''mass'', M,  to refer the NS gravitational mass, given by
\begin{equation}
\label{mgra}
M=M_G = \int_0^R 4 \pi r^{2} \, \epsilon(r) \, \mathrm{d}r \,. 
\end{equation}
From here on we will use $M_G$ for the gravitational mass to distinguish it from the baryonic mass, $M_B$, given by
\begin{equation}
\label{mbar}
M_B = m_N \int_0^R \frac{4 \pi r^{2} \, n_B(r)}{[1 - 2 G m(r)/r]^{1/2}} \, \mathrm{d}r \,, 
\end{equation}
 where $m_N$ is the nucleon mass and $n_B(r)$ is the baryonic number density.
 
The study of the evolutionary process of proto-NS in a $M_G-M_B$ plane was developed before by \cite{Bombaci:1996}. He has pointed out that the concept of NS maximum mass introduced by Tolmann, Oppenheimer and Volkoff \citep{tov1,  tov2}, $M_{TOV}$, defined as the maximum gravitational mass of a neutron star before becoming unstable and collapsing into a black hole, is partially inadequate when the first few seconds of NS evolution process are considered. He proposed that the study of gravitational mass - baryonic mass plane is the best way to analyze the stability of proto-NS and the maximum mass concept. In order to apply this study to proto-HS, we consider that the baryonic mass remains constant during the three stages of the evolution process because most of the matter accretion occurs in previous stages of the proto-NS \citep{Chevalier:1989}.

\begin{figure}[tb]
  \centering
  \includegraphics[width=0.95\columnwidth]{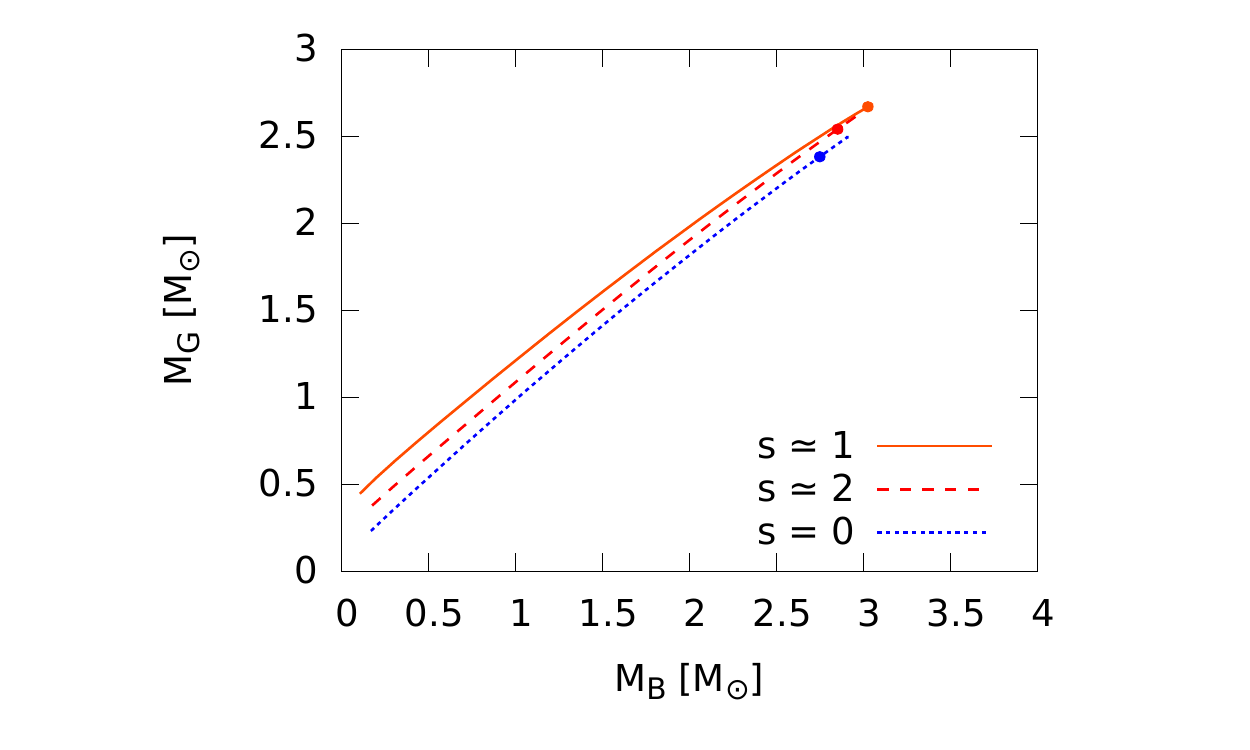}
  \caption{$M_{G} - M_{B}$ plane for Set 1 ($V_1 = \unit[0.02] {GeV}$, $G_2 = \unit[0.016] {GeV}^4$). In each constant-entropy curve, the rounded dot indicates the appearance of the quark matter core.}
  \label{mbmg1}%
\end{figure}

\begin{figure}[tb]
  \centering
  \includegraphics[width=0.95\columnwidth]{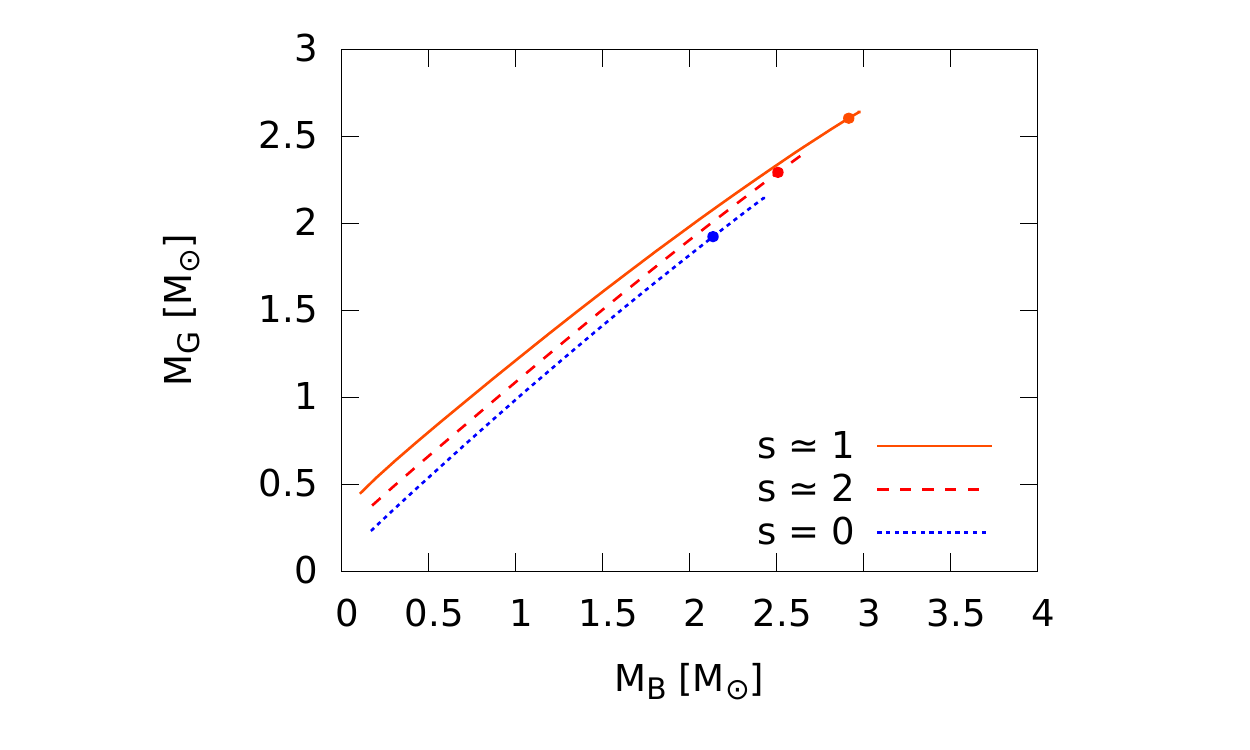}
  \caption{$M_{G} - M_{B}$ plane for Set 2 ($V_1 = \unit[0.03] {GeV}$, $G_2 = \unit[0.010] {GeV}^4$). In each constant-entropy curve, the rounded dot indicates the appearance of the quark matter core.}
  \label{mbmg2}%
\end{figure}

\begin{figure}[tb]
  \centering
  \includegraphics[width=0.95\columnwidth]{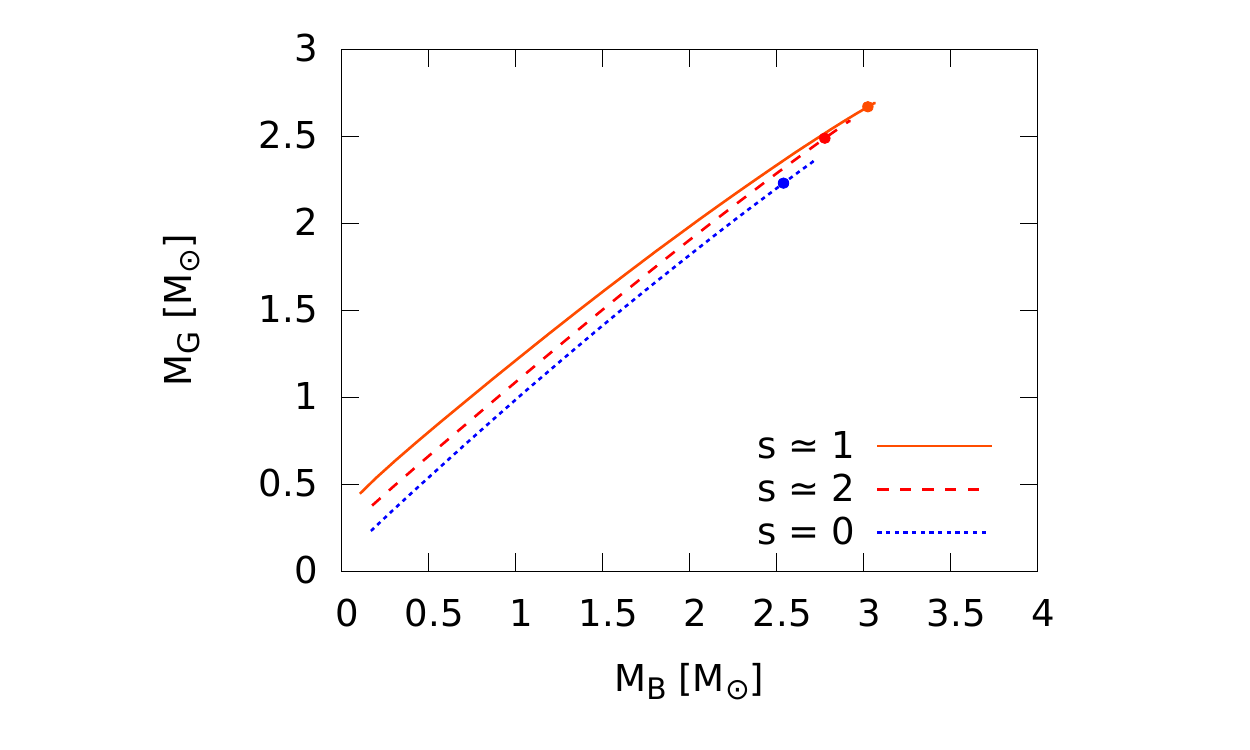}
  \caption{$M_{G} - M_{B}$ plane for Set 3 ($V_1 = \unit[0.10] {GeV}$, $G_2 = \unit[0.007] {GeV}^4$). In each constant-entropy curve, the rounded dot indicates the appearance of the quark matter core.}
  \label{mbmg3}%
\end{figure}

The results of the $M_G-M_B$ plane for the three sets of parameters considered  in this work are shown in Fig. \ref{mbmg1} (Set 1), Fig. \ref{mbmg2} (Set 2), and  Fig. \ref{mbmg3} (Set 3).  Each plane contains the three isentropic stages, $s \simeq 1$, $s \simeq 2$ and $s = 0$, taken into account for the evolution of the proto-HS. It can be seen that  $M_G$ increases with  $M_B$ up to a maximum value, after which the solutions become unstable. These unstable branches have been removed in all the plots to make the figures more readable.

\begin{figure}[tb]
  \centering
  \includegraphics[width=0.95\columnwidth]{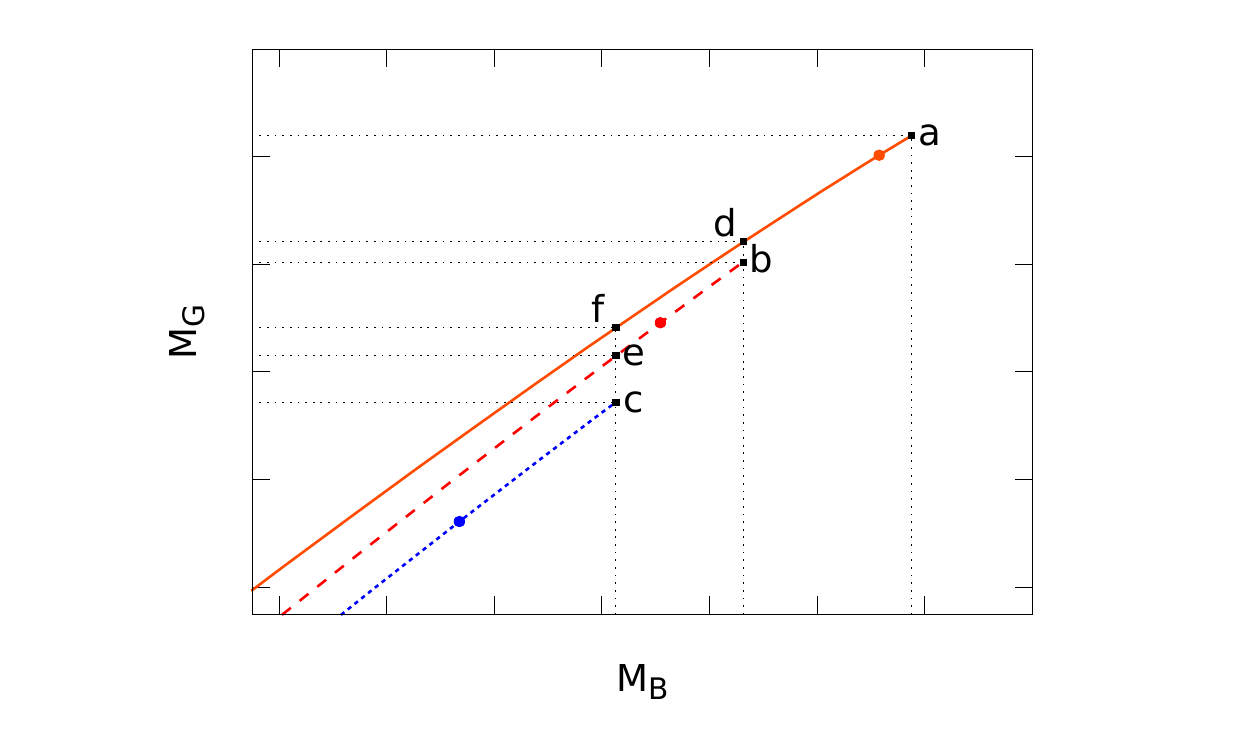}
  \caption{Enlarged $M_{G}-M_{B}$  plane for $V_1 = \unit[0.03] {GeV}$ and $G_2 = \unit[0.010] {GeV^4}$ close to the maximum mass configurations. In each constant-entropy curve, the rounded dot indicates the appearance of the quark matter core. The squared points \textit{a}, \textit{b} and \textit{c} indicate the maximum mass configuration for  $s \simeq 1$, $s \simeq 2$ and $s = 0$, respectively. The squared dot \textit{d} indicates the initial configuration that evolves, through a constant baryonic mass vertical line, to the maximum mass configuration of the curve $s \simeq 2$. Only for proto-hybrid stars with baryonic masses $0  <  M_B \leq M_B^{(f)}$, a constant baryonic mass vertical line can be drawn across the dots \textit{f}, \textit{e} and \textit{c} and the three isentropic curves. Thus, such stars will remain stable, surviving the evolutionary process towards a cold HS.}
  \label{mbmg4}%
\end{figure}

Following the study of \cite{Bombaci:1996}, we analyze qualitatively the evolution of proto-HS through Fig. \ref{mbmg4} considering the baryonic mass remains constant during the thermal evolution of the HS. We use this figure in order to clarify the evolution of HS from the $s \simeq 1$ to the $s = 0$ stages because it is representative of the behavior of $M_G-M_B$ curves shown in the Figs. \ref{mbmg1}, \ref{mbmg2} and \ref{mbmg3}.  The dots labeled by \textit{a}, \textit{b}, \textit{c} represent the maximum mass configuration corresponding to the isentropic curves $s \simeq 1$, $s \simeq 2$ and $s = 0$, respectively. Also, it is convenient to define $M_{B,G}^{(i)}$ as the baryonic/gravitational mass associated to \textit{i = a, b, c, d,} and \textit{f} in the Fig. \ref{mbmg4}.

We will also define four baryonic mass ranges:  $(0-M_B^{(f)}]$,  $(M_B^{(f)}-M_B^{(d)}]$, $(M_B^{(d)}-M_B^{(a)}]$ and $(M_B^{(a)}-\infty)$. The evolutionary process of the proto-hybrid stars can be followed starting from the $s \simeq 1$ curve, downwards along a constant baryonic mass vertical line. If $0  <  M_B \leq M_B^{(f)}$, as the vertical line crosses the $s \simeq 2$ and $s = 0$ curves, the proto-NS will remain stable, the neutrinos will escape and it will end as a cold neutron star in the $s = 0$ curve. If $M_B^{(f)} < M_B \leq M_B^{(d)}$ the proto-NS will reach a stable configuration in the $s \simeq 2$ curve and the neutrinos will escape, but then it will collapse to a black hole, meaning that no cold stable configurations are possible for these values of baryonic mass. For $M_B^{(d)} < M_B \leq M_B^{(a)}$ , the proto-NS newly formed will collapse shortly afterwards to a black hole. Finally, for  $M_B^{(a)} < M_B < \infty$, the proto-NS will never form and the supernovae remnant will collapse to a black hole.

The possibility of a phase transition in the proto-NS is indicated by rounded dots in the $M_{G}-M_{B}$ plane, suggesting the appearance of a quark matter core. Configurations with smaller baryonic masses than the associated to the rounded dot in each curve ($M_B^{(q,\,s=i)}$, $i=1,2,0$) are purely hadronic, while configurations with greater baryonic masses are neutron stars with a quark matter core. From Figs. \ref{mbmg1}, \ref{mbmg2} and \ref{mbmg3}, it can be seen that quark matter could be formed during the constant baryonic mass evolution. Thus, a proto-NS having a baryonic mass $M_B^{(q,\,s=0)} < M_B < M_B^{(f)}$ ending as a stable cold NS, would begin as purely hadronic and become an HS with a quark matter core after the whole cooling process.

According to our results, the sequence of snapshots suggests that a quark matter core may be formed during the cooling process between the stages with entropies per baryon $s \simeq 2$ and $s=0$. This is in contrast with the predictions of earlier studies where the deconfinement of quark matter occurs some seconds after the core bounce \citep{Benvenuto:1989tf,Benvenuto:1989qr} or the transition from nuclear to quark matter takes place during the stage with $s \simeq 1$ \citep{Benvenuto:1999}.

The first order phase transition between hadronic and quark matter in the proto-NS is triggered by the nucleation of a critical size drop of the new stable quark phase in the metastable hadronic phase. It has been suggested in a detailed study  of the effects of quark matter nucleation on the evolution of proto-NS by \cite{bombaci:2011} and \cite{bombaci:2016ep} that if the nucleation time is greater than the cooling time for a proto-hadronic star, then quark matter nucleation is not likely to occur in a newly formed star. 

The calculation of the nucleation time depends on the properties at the interface between the hadronic and quark matter phases and it is out of the scope of this study. But based on the results of \cite{bombaci:2011} and \cite{bombaci:2016ep} , we assume that the nucleation time of quark matter is less than the cooling time of the proto-NS during its evolution when the entropy per baryon decreases from $s \simeq 2$ to $s=0$. In this way, is feasible the proto-NS evolves into a cold hybrid configuration with a quark matter core.

Our results show that it is always possible to obtain hydrostatic stable configurations for the different parameter combinations used. Also, for some particular combinations, the two solar masses for the gravitational maximum mass star are reached. It is worth noting that an increase in the value of $V_1$ or $G_2$ generates an increase in the value of this maximum mass. For CHS, $s = 0$, stable configurations may suggest that PSR J1614-2230 and PSR J0348+0432 stars could be HS. 

Although there is not still an accurate enough determination of NS radii, we have considered the constraint imposed by two recent studies on NS: a lower limit of  $R > \unit[10.7] {km}$ for NS with masses $M = \unit[1.4] {M_{\odot}}$ established by \cite{Chen:2015}, and an upper limit of $R < \unit[16.8] {km}$, inferred from the NS RX J1856-375 \cite{Boshkayev:2016}. The radii obtained for the CHS are within the range set by such limits.

%==========================================================================================================

\section{Summary and conclusions}
\label{conclus}

In this work we have studied, in a very simplified way, the thermal evolution of proto-NS modeled as HS, considering a successive sequence of constant entropy per baryon snapshots. The evolution started in a hot stage, with and entropy per baryon of $s \simeq 1$ and where trapped neutrinos have been subjected to the condition $Y_{L_e} = 0.4$. In the following stage, after $\sim 15$ seconds, neutrinos diffuse and warm the star being $s \simeq 2$ and $Y_{\nu_e} = 0$. Finally, after a few minutes, the star cools down and the last stage is reached, where $s = 0$.

We used a tabulated EoS to describe the hadronic phase of the proto-hybrid stars and the FCM for the description of quark matter. To analyze how the results for proto-hybrid stars evolution depend on the FCM parameters, we have taken into account three different combinations of $V_1$ and $G_2$.

The use of the NeStOR code to construct the hybrid EoS, allowed us to study the HS structure via the integration of the TOV equations as well as the gravitational mass - baryonic mass plane, $M_G-M_B$. The hybrid EoS was constructed at finite temperature, taking into account the characteristic composition for each evolutionary stage. For the hadronic phase, we used an EoS constructed for astrophysical simulation of proto-NS. We have obtained, for each constant entropy stage, a family of stable HS, with a characteristic mass-radius relationship. Considering the constraints from astrophysical observations of NS, we have studied the variation of the maximum mass star with the different combination of FCM parameters, for each family. 

Also in the framework of the FCM semi-analytical quark matter EoS, we analyzed the stability of strange quark matter within the strange quark matter hypothesis. The results obtained from our semi-analytical quark EoS are in agreement with previous works using the FCM, both the simplified phase diagram of the QCD as the stability of the strange quark matter. Under the simplified model of thermal evolution considered in this work, the temperature profiles obtained for the HHS are in agreement with the dynamical calculations with schematic EoS of the thermal evolution of NS.

Regarding the concept of maximum mass for proto-NS, we have also analyzed the $M_G-M_B$ plane. The cooling evolution was studied considering the baryonic mass conservation during the whole process \citep{Bombaci:1996}. We have schematically analyzed the baryonic mass ranges, separating the newly born NS, which will evolve to stable CHS from those that will collapse afterwards to a black hole.

The study of the gravitational mass-baryonic mass plane for proto-HS, has revealed in which of the three stages of constant entropy considered, a phase transition from hadronic matter to quark matter could occur in the context of the microscopic models used in this work. Our results suggest that an early ($s \simeq 1$) phase transition between quark matter and hadronic matter is not produced for the baryonic mass ranges in which the proto-HS end as stable CHS, but the formation of stable CHS is possible due a late phase transition, after the stage $s \simeq 2$. This contrasts with previous studies \citep{Benvenuto:1989tf, Benvenuto:1989qr, Benvenuto:1999} where pure hadronic stars are converted to quark stars within the first seconds after their birth.

It is worth pointing out that our model is a simplified treatment of the thermal evolution. The calculation of the quark matter nucleation rate and the nucleation time, due to thermal nucleation mechanisms in the framework of the EoS considered, would provide a better understanding of the finite size effects of the phase transition between hadronic and quark matter. These further calculations are beyond the scope of this work.

We have obtained stable CHS that easily reach the two solar masses for the maximum mass star for several combination of the FCM parameters. According our results, the pulsars J1614-2230 and J0348+0432 would contain quark matter in their inner cores. However, we note that we have neither considered hyperons in the hadronic phase nor the color superconductivity in the quark-gluon plasma phase. The inclusion of either of these two contributions will modify the EoS of each phase and therefore the critical point where the transition occurs. The effects of such contributions should be studied in a future work.

Future observations of neutrinos in supernovae, as well as the observation of $x-ray$ and $\gamma-ray$ emissions of young pulsars will enable the astrophysicist community to carry out simulations on real thermal evolution that will contribute to the understanding of the underlying physics of NS.

%==========================================================================================================

\begin{acknowledgements}
       The authors thank the referee for the constructive comments and criticisms that have contributed to improve the manuscript substantially.  M. M. and  M. O. thank CONICET and UNLP for financial support. M. O. thanks the financial support from  American Physical Society's International Research Travel Award Program and National Science Foundation under grant no. PHY-1411708. We thank G. A. Contrera for his help with some of the figures.
\end{acknowledgements}


\bibliographystyle{aa}
\bibliography{biblio_mariani}

\begin{appendix} %First appendix
\section{Multi expansion}
\label{apendice1}

For the quark pressure within the FCM, we
made two series expansions to calculate the Fermi integrals. This
approach will allow us to work at finite temperature, T, and chemical potential, $\mu$, regime.

The pressure expression for a single quark within the FCM \citep{Simonov:2007a,Simonov:2007b,Nefediev:2009} is given by
\begin{equation}
	 P_q = \frac{T^4}{\pi^2} \int_0^\infty du \frac{u^4}{\sqrt{u^2+(\frac{m_q}{T})^2}}\frac{1}{\exp{\Big[\sqrt{u^2 + (\frac{m_q}{T})^2} - \frac{\mu_q - V_1/2}{T}\Big]} + 1}\, .
\end{equation}
By defining  $u=p/T$ and $\epsilon=(p^2+m_q)^{1/2}+\frac{V_1}{2}$, we obtain
\begin{equation}
 P_q = \frac{1}{\pi^2}\int_{m_q}^\infty d\epsilon [(\epsilon-\frac{V_1}{2})^2-m_q^2]^{3/2} \frac{1}{e^{(\epsilon-\mu)/T}+1} \, .
\end{equation}
After a change of variables,
\begin{equation}
  \tilde{\epsilon}=\epsilon-\frac{V_1}{2} \, ,
\end{equation}
\begin{equation}
  \tilde{\mu}=\mu-\frac{V_1}{2} \, ,
\end{equation}
we get
\begin{equation}
 P_q = \frac{1}{\pi^2}\int_{m_q}^\infty d\epsilon [\tilde{\epsilon}^2-m_q^2]^{3/2} \frac{1}{e^{(\tilde{\epsilon}-\tilde{\mu})/T}+1} \, ,
\end{equation}
which can be expand in power series of  $m_q^2/\tilde{\epsilon}^2$ up to second order. This expansion is based on the work of  \cite{Masperi:2004} and allow us to obtain an analytic approximation  given by
\begin{equation}
\label{pres}
\,P_q\simeq \frac{T^4}{\pi^2} \exp x\int_{\exp (a)}^{\infty}\frac{d\eta }{\eta \left( \eta +\exp x\right) }\left( \ln ^{3}\eta -\frac{3}{2} a^{2}\ln \eta +\frac{3}{8}\,\frac{a^{4}}{\ln \eta}\right) \, ,
\end{equation}
where  $a=m/T$ and $x=\tilde{\mu}/T$.

To integrate Eq.(\ref{pres}) we have to distinguish two cases. First, if $x<a$ $(\tilde{\mu}<m_q \rightarrow \mu-V_1/2<m_q)$, we expand the expression in terms of $\exp(x)/\eta$ up to fourth order obtaining
  
  \begin{eqnarray}\label{desarrollo1}
 P_q &\simeq & \frac{T^4}{\pi^2} \big[\exp \left( x-a\right) \;\left( 6+6a+\frac{3}{2}a^{2}-\frac{a^{3}}{2}\right) + \nonumber \\ 
&& -\frac{\exp \left[ 2\left(x-a\right) \right] }{2}\left(\frac{3}{4}+\frac{3}{2}a+\frac{3}{4}a^{2}-\frac{a^{3}}{2}\right) + \nonumber \\
&& +\frac{\exp \left[ 3\left( x-a\right) \right] }{3}\left(\frac{2}{9}+\frac{2}{3}a+\frac{a^{2}}{2}-\frac{a^{3}}{2}\right) - \nonumber \\
&& -\frac{3}{8}a^{4} \big[ \exp(x) Ei\left( -a\right) -\exp \left(2x\right) Ei(-2a)+ \nonumber \\
&& +\exp \left( 3x\right) Ei\left( -3a\right) \big]\big]\;.
  \end{eqnarray}
  
Second, if $x>a$ $(\tilde{\mu}>m_q \rightarrow \mu-V_1/2>m_q)$, we divide the integral into two parts,
\begin{equation}
	\label{disp}
  	\int_{\exp (a)}^\infty [...] \, d\eta \, = \int_{\exp (a)}^{\exp(x)} [...] \, d\eta \, + \int_{\exp (x)}^\infty [...] \, d\eta \, .
\end{equation}
The result of the second integral in the right hand side of Eq. (\ref{disp}) is identical to the previous expansion, Eq.(\ref{desarrollo1}), in terms of $\exp(x)/\eta$. For the first integral, we make a Laurent expansion in terms of $\eta/\exp(x)$. The final result of both expansions gives

 \begin{eqnarray}\label{desarrollo2}
P_q &\simeq& \frac{T^{4}}{\pi ^{2}}  \big[\frac{x^{4}}{4}+\frac{x^{3}}{3}+\frac{29}{6} x^{2}+\frac{2}{9}x+\frac{1223}{108}+\frac{a^{4}}{2}+ \nonumber \\ 
&& -\frac{3}{2}a^{2}\left(\frac{x^{2}}{2}+\frac{x}{3}+\frac{29}{18}\right) + \nonumber \\ 
&& +\exp \left[-\left( x-a\right) \right] \left(-6+6a-\frac{3}{2}a^{2}-\frac{a^{3}}{2}\right) + \nonumber \\ 
&& +\frac{\exp \left[ -2\left( x-a\right) \right] }{2}\left( \frac{3}{4}- \frac{3}{2}a+\frac{3}{4}a^{2}+\frac{a^{3}}{2}\right) + \nonumber \\
&& +\frac{3}{8}a^{4}[-\exp(x)\,Ei\left( -x\right) + \exp(2x)\,Ei(-2x) + \nonumber \\
&& -\exp \left( 3x\right) \,Ei\left(-3x\right) - \exp \left( -x\right) \,Ei\left( x\right) + \nonumber \\
&& +\exp \left(-2x\right) \,Ei\left( 2x\right) +\exp \left( -x\right) \,Ei\left(a\right) - \nonumber \\ 
&& -\exp \left( -2x\right) \,Ei\left( 2a\right) +\ln x-\ln a] \big]\;.
  \end{eqnarray}

\end{appendix}

\end{document}